\documentclass[12pt, draftclsnofoot, onecolumn]{IEEEtran}
\IEEEoverridecommandlockouts

\usepackage[font=footnotesize,caption=false]{subfig} 

\usepackage{cite}
\usepackage{amsmath,amssymb,amsfonts, mathtools, amsthm}
\usepackage{algorithmic}
\usepackage{graphicx}
\usepackage{textcomp}
\usepackage{xcolor}
\usepackage{booktabs}
\usepackage{glossaries}
\usepackage{subfig}
\usepackage{algorithm}
\usepackage{algorithmic}
\usepackage{bbm}
\usepackage{bm}
\usepackage{multirow}
\newcommand{\set}{\mathcal}
\usepackage[normalem]{ulem}

\newtheorem{theorem}{Theorem}

\newacronym{snr}{SNR}{signal-to-noise ratio}
\newacronym{bcd}{BCD}{Block Coordinate Descent}
\newacronym{leo}{LEO}{Low Earth Orbit}
\newacronym{isl}{ISL}{Inter-Satellite Link}
\newacronym{gsd}{GSD}{ground sample distance}
\newacronym{fov}{FoV}{field of view}
\newacronym{gtfp}{GTFP}{ground track frame period}
\newacronym{gs}{GS}{ground station}
\newacronym{fso}{FSO}{free-space optical}
\newacronym{smec}{SMEC}{satellite mobile edge computing}
\newacronym{mec}{MEC}{mobile edge computing}
\newacronym{ipdd}{IPDD}{Increasing Penalty Dual Decomposition}

\definecolor{BuGn}{RGB}{28,144,153}

\newcommand{\nwtxt}[1]{#1}

\DeclareMathOperator*{\maximize}{maximize}
\DeclareMathOperator*{\minimize}{minimize}

\DeclareMathOperator*{\argmin}{arg\,min}

\def\BibTeX{{\rm B\kern-.05em{\sc i\kern-.025em b}\kern-.08em
    T\kern-.1667em\lower.7ex\hbox{E}\kern-.125emX}}

\newcounter{storeeqcounter}
\newcounter{tempeqcounter}

\title{Satellite edge computing for real-time and very-high resolution Earth observation}
\author{Israel Leyva-Mayorga, \IEEEmembership{Member,~IEEE}, 
Marc M. Gost,\\
Marco Moretti, \IEEEmembership{Member,~IEEE} 
Ana Pérez-Neira, \IEEEmembership{Fellow,~IEEE},\\ 
Miguel Ángel Vázquez, \IEEEmembership{Senior Member,~IEEE}
Petar~Popovski, \IEEEmembership{Fellow,~IEEE},\\
and Beatriz~Soret, \IEEEmembership{Senior Member,~IEEE}
\thanks{Israel Leyva-Mayorga and Petar Popovski are with the Department of Electronic
Systems, Aalborg University, 9220 Aalborg, Denmark (e-mail:\{ilm, petarp\}@es.aau.dk). Marc M. Gost is with the Centre Tecnològic de Telecomunicacions de Catalunya and the Dept. of Signal Theory and Communications, Universitat Politècnica de Catalunya, Spain (email: marc.martinez.gost@upc.edu). Marco Moretti is with the Dept. of Information Engineering, University of Pisa, Italy (email: marco.moretti@unipi.it). Ana Pérez-Neira is with ICREA, the Centre Tecnològic de Telecomunicacions de Catalunya and the Dept. of Signal Theory and Communications, Universitat Politècnica de Catalunya, Spain (email: ana.perez@cttc.es). Miguel Ángel Vázquez is with the Centre Tecnològic de Telecomunicacions de Catalunya, Spain (email: mavazquez@cttc.cat). 
Beatriz Soret is with the Telecommunications Research Institute, University of Málaga, Spain and with the Department of Electronic
Systems, Aalborg University, 9220 Aalborg, Denmark (email: bsoret@ic.uma.es).} \thanks{This work was partially funded by SatNEx-V, co-funded by the European Space Agency (ESA). The work of Ana P\'erez-Neira is partially supported by Project IRENE- (PID2020-115323RB-C31) funded by MCIN/AEI/ 10.13039/501100011033.}}
\date{}

\begin{document}

\maketitle
\vspace{-2em}
\begin{abstract}
\nwtxt{In high-resolution Earth observation imagery, \gls{leo} satellites capture and transmit images to ground to create an updated map of an area of interest. Such maps provide valuable information for meteorology and environmental monitoring, but can also be employed for real-time disaster detection and management. However, the amount of data generated by these applications can easily exceed the communication capabilities of \gls{leo} satellites, leading to congestion and packet dropping. To avoid these problems, the \glspl{isl} can be used to distribute the data among multiple satellites and speed up processing. In this paper, we formulate a \gls{smec} framework for real-time and very-high resolution Earth observation and optimize the image distribution and compression parameters to minimize energy consumption. Our results show that our approach increases the amount of images that the system can support by a factor of $12\times$ and $2\times$ when compared to directly downloading the data and to local \gls{smec}, respectively. Furthermore, energy consumption was reduced by $11$\% in a real-life scenario of imaging a volcanic island, while a sensitivity analysis of the image acquisition process demonstrates that energy consumption can be reduced by up to $90$\%.}

\end{abstract}
\glsresetall

\glsresetall
\section{Introduction}
Satellites in \gls{leo} are widely used for Earth observation purposes as they can construct high-resolution maps of large areas by capturing images from space as they orbit the Earth. These maps can be used in in various applications, as in meteorology, agriculture, or environmental monitoring~\cite{Herold2008}. They are also very valuable
in near-real time applications, such as disaster detection and identification, supporting the coordination of the  emergency response.
Due to the limited storage in the individual \gls{leo} satellites, the images must be 1) captured, 2) processed for compression and/or optical correction, and 3) transmitted to ground for storage and/or distribution across the terrestrial infrastructure.

The area covered by each individual satellite image depends on numerous factors. Namely, the \gls{fov} of the instrument/camera is the angle that determines the observable space of the camera sensor. The \gls{fov} together with the altitude of deployment of the satellite determine the \emph{swath}, which is the width of the observable area on the surface of the Earth. Then, the swath and the resolution of the instrument/camera determine the \gls{gsd}, which is the distance covered by each single pixel. As an example, the European Space Agency (ESA) Sentinel-2 mission is located at an altitude of $748$\,km and captures images in the visible spectrum with a \gls{gsd} of $1.5$ m~\cite{Drusch2012}. In total, the swath for Sentinel-2's visible spectrum instrument is only $60$\,km.
Other common values for the \gls{gsd} are in the range between $0.3$ and $3$ meters~\cite{Denby2019, Denby2020}.  

To achieve a high image resolution, the \gls{gsd} must be small and, in turn, so does the area covered by a single image. 
For example, the area covered by an image in high-definition (HD) format with a \gls{gsd} of 3\,m is around $18\,\text{km}^2$, which is smaller than several European airports. 
Therefore, high-resolution maps of large areas must be created by capturing and organizing a large number of images. These images must be captured with a sufficiently high frequency to avoid coverage holes in the map.

Traditionally, the Earth observation satellites would communicate directly with a \gls{gs} to download the collected data. However, this greatly limits the amount of data that can be collected. In contrast, modern Earth observation missions possess numerous satellites with advanced communications and processing capabilities. For instance, recent advances in \gls{fso} communications technology~\cite{Kaushal2017}, in combination with the stable relative positions of satellites in the same orbital plane, make it possible to establish intra-plane \glspl{isl} to connect neighboring satellites in the same orbital plane using high-data rate \gls{fso} links. Such links can be used to distribute the images across the neighboring satellites, forming a \gls{smec} cluster with distributed processing capabilities.

\begin{figure}[t]
\centering
\subfloat[]{\includegraphics{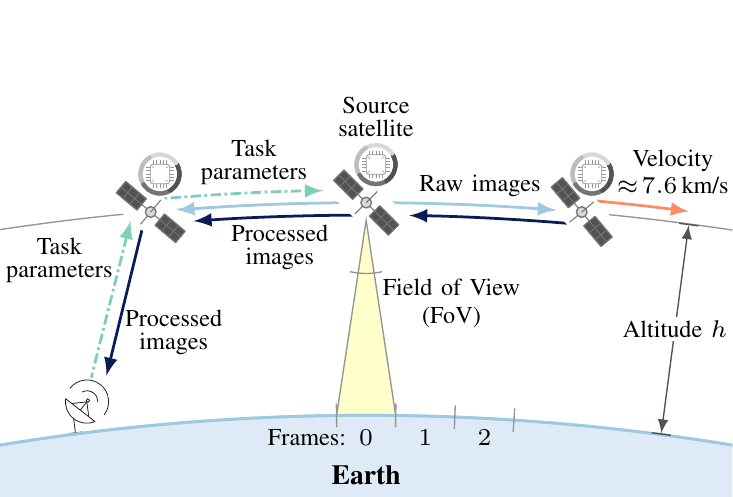}\label{fig:sat_fov}}\hfill
\subfloat[]{\includegraphics{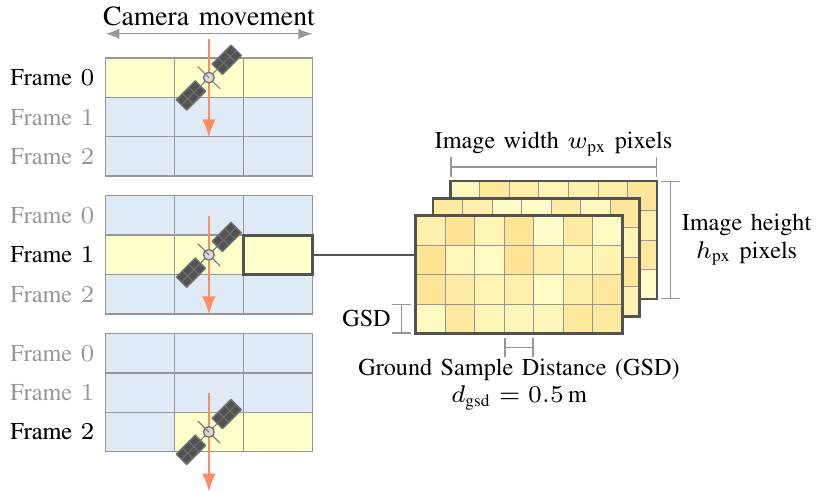}\label{fig:top_view}}
\caption{Earth observation application where \nwtxt{the source satellite} scans the area of interest. (a) The \gls{fov} and the altitude of the satellites determine the area covered by each image. The raw images \nwtxt{captured by the source satellite, in the middle of the figure,} can be shared with the nearby satellites to process them in a distributed manner and, then, send them to the \gls{gs}.
(b) The \nwtxt{source} satellite captures a frame, consisting of $W_k$ images, at slot $k$ as indicated by the yellow area. The value of $W_k$ may be different for each $k$ depending on the dimensions of the area of interest. }
\label{fig:fov_diag}
\end{figure}

Fig.~\ref{fig:fov_diag} illustrates an Earth observation mission relying on a \gls{smec} cluster to distribute the data to be processed and compressed by multiple satellites  before being downloaded to the \gls{gs}. As it can be seen, the orbital velocity of the satellite creates the need to capture images frequently to avoid coverage holes and the frequency is determined by the \gls{fov} and the altitude of the satellite. Specifically, the maximum period at which the images must be captured to avoid coverage holes is called the \gls{gtfp}, which imposes the timing constraints in the satellite processing and communication subsystems. In such setup, illustrated in Fig.~\ref{fig:top_view}, the system operation can be divided into time slots. At each time slot, the satellite scans the area of interest by capturing a set of side-by-side images called a frame. Then, to maintain the stability in the processing and communication subsystems at the satellites, the time needed for processing and for communication at each link must be lower than that of the duration of a time slot. This is the scenario of interest for the present paper.

The generation of enormous amounts of data in Earth observation applications is a common concern, especially regarding the downlink capacity (i.e., satellite-to-ground link). \nwtxt{Moreover, as satellites rely on solar panels and rechargeable batteries for energy supply, minimizing the energy consumption of the communication and processing task at the satellites is of utmost importance. Nevertheless, the research on \gls{smec} to reduce the amount of data transmission and energy consumption in Earth observation applications is still in its infancy. Specifically, most of the literature on \gls{smec}, with a few clear exceptions~\cite{Denby2019,Denby2020} that include our previous work~\cite{marc2022}, deals with tasks generated at the mobile users and, hence, mimics the traditional functionality of a terrestrial \gls{mec} but substituting the terrestrial edge nodes with satellites~\cite{Wei2019}. Currently, \gls{smec} approaches for Earth observation can be divided into: early discard~\cite{Denby2019, Denby2020} and compression~\cite{marc2022}}.  Early discard of images with relatively low information content (e.g., cloud coverage) effectively reduces the amount of data transmitted to ground~\cite{Denby2019, Denby2020}. Note that the efficiency of early discard is mostly based on autonomous decisions at the satellites, which determine which images to transmit to ground and which ones to discard. Even though this inference process can be enhanced with techniques such as image-chain simulation (ICS) to evaluate the quality of the captured images~\cite{Alici2019}, it may be problematic for mission-critical applications, for example, emergency scenarios, where partially obstructed images can still be valuable for the \gls{gs}. To avoid these problems, the satellites might instead execute an algorithm to compress the collected data. \nwtxt{For example, new video coding implementations for satellite Earth observation focus on achieving a sufficiently high encoding rate and compression ratio so the video frames can be compressed and transmitted to the \gls{gs} based on the limitations of the downlink and processing modules~\cite{bui2022videoencoding}. The authors in~\cite{bui2022videoencoding} focus on the algorithmic implementations in a scenario with a single satellite and fixed downlink data rate and do not optimize the coding process for energy efficiency. Even though the proposed algorithm improved the execution time of the video coding process, it might still be restrictive for video applications requiring a high frame rate video. Thus, this scenario presents a good candidate for the distribution and parallel processing of the video frames at multiple satellites as proposed in the present paper.}

While the literature on \gls{smec} for Earth observation focuses on local processing of the data, there are numerous examples of terrestrial \gls{mec} and \gls{smec} that consider the optimization of distributed processing, which is a generalization of local processing. For example, lossless compression was considered to occur both at the source mobile device and at the edge server in a terrestrial \gls{mec} to reach a combined compression ratio~\cite{Wang2020}. In a similar fashion, the benefits of offloading computation tasks from mobile users in remote areas to a \gls{smec} network were studied in \cite{tang2021computation}, which builds on recent literature on \gls{smec} architectures \cite{qiu2019deep,cheng2019space,alsharoa2020improvement,liu2020task}. In~\cite{tang2021computation}, authors propose a three-tier scheme where the  tasks with low computational complexity are directly processed at the mobile users and the rest are distributed either to a ground cloud or to the satellite edge. The complex allocation problem is solved with the alternating direction method of multipliers (ADMM). 

 In our previous work~\cite{marc2022}, we focused on an Earth observation scenario with power allocation for communication and where the source satellite chooses between directly downloading the data or compressing it locally.
\nwtxt{The present paper represents a major extension of our previous work and a major deviation with respect to the \gls{smec} literature described above by considering real-time high-resolution Earth observation applications where the real-time requirements are defined by the dynamics of the system. In the considered scenario, we consider a general \gls{smec} framework for real-time and very-high resolution Earth observation that involves five phases: 1) segmentation of the imaging data, 2) allocation of the processing tasks, 3) distribution of the image segments (i.e, scatter), 4) processing of the image segments, and 5) delivery of the processed images to the ground station (i.e., gather). }

\begin{figure}[t]
    \centering
    \includegraphics{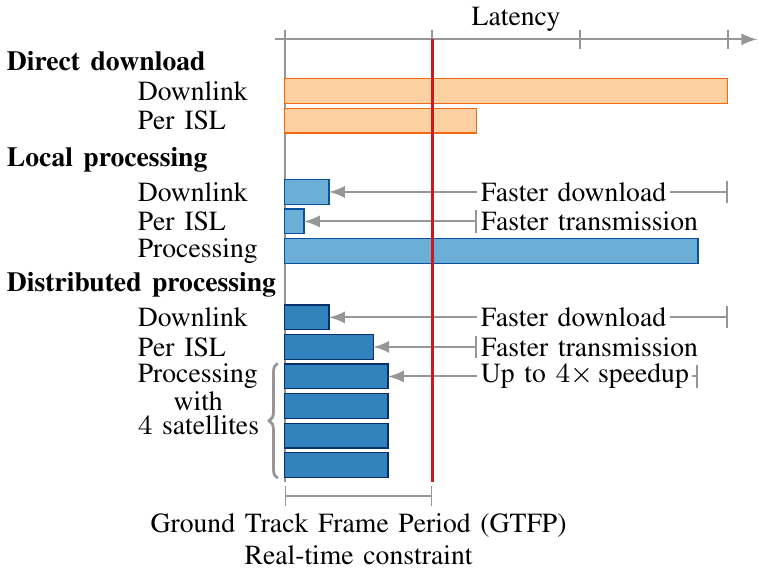}
    \caption{Illustration of the different options for data delivery in high-resolution Earth observation: direct download, local processing, and distributed processing.}
    \label{fig:diags}
\end{figure}

 Fig.~\ref{fig:diags} illustrates the diverse options for executing a task with the \gls{gtfp} as real-time constraint. \nwtxt{The data can be directly downloaded to the \gls{gs}. However, this might may cause backlog in the satellite-to-\gls{gs} and \gls{isl} links due to their limited data rate for communication. On the other hand, the data can be processed locally at the source satellite. However, this might cause backlog at the CPU of the source satellite due to the limited processing power. Finally, the data can be distributed across $4$ satellites, which alleviates the load in both the communication links and at the individual CPUs. This would reduce the amount of data to be transmitted at each individual link and to be processed at each CPU. By doing so, the real-time constraint defined by the \gls{gtfp} is fulfilled, which guarantees the stability of the communication and processing subsystems.}

The main contributions of this paper are described in the following.
\begin{enumerate}
\item We define a novel and general model for high-resolution Earth observation imagery, along with its real-time constraints imposed by the physical (i.e., orbital) parameters of \gls{leo} constellations and of the imaging instruments.


\item We formulate and solve a global optimization problem for the segmentation, allocation, and processing phases of our general \gls{smec} framework. Given that the satellites have a limited battery supply, the objective of distributing the tasks across the satellites in the constellation is to minimize the overall energy consumption while fulfilling the limitations of the processing frequency at the satellites' CPU, and the rates at the  \glspl{isl} and satellite-to-ground link. \nwtxt{The present work is a major extension of our previous study~\cite{marc2022}, where we only considered local edge computing and cloud computing, a single and homogeneous task at the satellites, a fixed CPU frequency, and a fixed compression ratio. In the present paper, we consider the optimization with distributed processing, where the set of satellites participating in the processing, the CPU frequency, and the compression ratio are optimization variables. Hence, the optimization problem considered in this paper is a generalization of the one considered in our previous work~\cite{marc2022}.}

\item \nwtxt{We consider a realistic imaging data size and perform both 1) a sensitivity analysis of the impact of the data generation process on the optimal solution of the problem and 2) a case study that validates our approach and illustrates of the potential gains of the global optimal solution in a real-life scenario.}
\end{enumerate}
\nwtxt{In the baseline scenarios, our results show that the number of images that can be transmitted to the \gls{gs} with our \gls{smec} approach is up to $12\times$ larger when compared to direct download without processing and up to $2\times$ larger when compared to processing at the source satellite. Furthermore, the energy consumption of our \gls{smec} approach is up to $10\times$ lower than with direct download. Finally, when compared to optimizing each frame independently, jointly optimizing the task parameters for the whole task with $K$ frames leads to energy savings of up to $90$\% in the baseline scenarios with a relatively low number of images per frame. In the real-life scenario, where  a satellite scans the island of La Palma, our results show that the energy consumption can be reduced at up to $11$\% by optimizing the task parameters for the $K=82$ frames when compared to optimizing for each frame independently This scenario is only feasible with distributed \gls{smec}, as both direct download without processing and local processing approaches would lead to buffer overflows in the communication and/or processing buffers of the satellites.}

The rest of the paper is organized as follows. Section~\ref{sec:system_model} describes the system model. Next, the optimization problem is formulated in Section~\ref{sec:problem}, numerical results are presented in Section~\ref{sec:results}, and concluding remarks are given in Section~\ref{sec:conclusions}.


\section{System model}
\label{sec:system_model}


\subsection{Framing}
We consider an orbital plane (i.e., ring) in a \gls{leo} satellite constellation and a \gls{gs} $g$. The orbital ring is comprised of $N$ satellites uniformly spaced along the orbit and deployed at an altitude $h$\,km above the Earth's surface. The satellites possess, among other subsystems, a camera, a processing payload, and communication modules to establish a radio frequency link towards the \gls{gs} as well as high-data rate \gls{fso} \glspl{isl} with the two neighboring satellites. Hence, the satellites in the orbital plane form a ring topology.

The satellites in the orbital plane serve an Earth observation application with pre-planned tasks. Each task involves capturing images to scan a specific area of interest, which must be transmitted to the \gls{gs}.
\begin{table}[t]
\centering
\caption{Parameters defined in the system model}
\begin{tabular}{@{}lll@{}}
\toprule
\multicolumn{2}{@{}l}{Symbol} & Description \\\midrule
\multicolumn{3}{@{}l}{\textbf{Scenario}}\\
&$N$ & Number of satellites in the orbital plane\\
&$h$ & Altitude of deployment of the satellites $[\mathrm{m}]$\\
&$h_\text{px}$ & Height of each image $[\mathrm{pixels}]$\\
&$w_\text{px}$ & Width of each image $[\mathrm{pixels}]$\\
&$q_\text{px}$ & Amount of bits per pixel\\
&$D_\text{img}$ & Size of each image $[\mathrm{bits}]$\\
&$W_k$ & Number of side-by-side images in frame $k$\\
&$d_\text{gsd}$ & Average ground sampling distance $[\mathrm{m}]$\\
&$T_\text{GTF}(h_\text{px},d_\text{gsd},h)$& Ground track frame period (GTFP) $[\mathrm{s}]$\\
\multicolumn{3}{@{}l}{\textbf{Processing}}\\
&$f_\text{CPU}$ & Highest CPU frequency $[\mathrm{Hz}]$\\
&$f_k^{(n)}$ & CPU frequency $[\mathrm{Hz}]$ for processing at satellite $n$ and frame $k$\\
&$N_\text{CPU}$ & Number of available processing cores per satellite\\
&$\rho_k$ & Compression ratio for frame $k$\\
& $\epsilon$ & Parameter that \nwtxt{determines} the complexity of the compression algorithm\\
& $C(\rho_k,\epsilon)$ & Number of CPU cycles to compress one bit of data\\
\multicolumn{3}{@{}l}{\textbf{Communication}}\\
&$R_k$ & Data rate to transmit frame $k$ to the \gls{gs} $[\mathrm{bps}]$ \\
&$P^\text{tx}_\text{RF}$ & \nwtxt{Transmission power for RF satellite-to-ground links} $[\mathrm{W}]$ \\
&$R_\text{ISL}$ & Fixed data rate at the ISLs $[\mathrm{bps}]$\\
&$P_\text{ISL}$ & \nwtxt{Power consumption of the ISLs during transmission} $[\mathrm{W}]$\\
&$\eta$ & \nwtxt{Fraction of $P_\text{ISL}$ consumed due to data transmission}\\
\multicolumn{3}{@{}l}{\textbf{Segment allocation}}\\
&$X\in\mathbb{Z}^{K,N}$ & Segment allocation matrix $[\mathrm{bits}]$\\
&$\mathbf{x}_k\in\{0,1,\dotsc,D_k\}^N$ & Segment allocation vector for frame $k$ $[\mathrm{bits}]$\\
&$x_k^{(n)}\in\{0,1,\dotsc,D_k\}$ & Segment allocation for satellite $n$ and frame $k$ $[\mathrm{bits}]$\\ 
     \bottomrule
\vspace{-1em}
\end{tabular}
\label{tab:system_params}
\end{table}
The camera is used to capture images of a fixed size \begin{equation}
    D_\text{img}=w_\text{px}\, h_\text{px}\, q_\text{px}\,\text{bits,}
\end{equation}
 where $w_\text{px}$ and $h_\text{px}$ are the width and height in pixels and $q_\text{px}$ is the number of bits to represent each pixel. Throughout this article, we assume that $h_\text{px}<w_\text{px}$ is aligned with the velocity vector of the satellite (i.e., the roll axis). The satellites capture images by pointing directly towards the Earth and by rotating the camera perpendicularly to the velocity vector as illustrated in Fig.~\ref{fig:top_view}. Therefore, the satellite itself must enter the area of interest to start capturing the images. 
 
 The average distance between adjacent pixel centers $d_\text{gsd}$ taken when pointing directly at the nadir point is called the \gls{gsd}, which is a function of $w_\text{px}$, the \gls{fov}, and the satellite altitude $h$~\cite{Denby2019}. The area covered by a single image is 
\begin{equation}
    A  \approx w_\text{px}h_\text{px}d_\text{gsd}^2
\end{equation}
and the vertical distance--aligned with the roll axis of the satellite--covered is $h_\text{px}d_\text{gsd}$\,m. 

The task and, similarly, the operation of the system are divided into $K$ time slots of duration equal to the \gls{gtfp}. At each time slot $k$, the source satellite $v_0$ captures a frame comprised of $W_k\in\mathbb{N}^+$ side-by-side images by rotating the camera perpendicularly to its velocity vector to cover the area of interest as it moves. The number of images $W_k$ can be different for each frame in a task as exemplified in Fig.~\ref{fig:top_view}, where frame $W_0=W_1=3$ images and $W_2=2$. 

We assume that the time to capture the $W_k$ images in a frame and to make them available for the processing algorithm and the communication modules (i.e., to write the data in memory) does not impact the real-time constraints of the system. The resulting size of the $k$-th frame is $D_k=W_kD_\text{img}$\,bits and the frame is approximately $W_kw_\text{px}d_\text{gsd}$\,meters wide.

Capturing frames at exactly the \gls{gtfp} ensures that there are no coverage gaps, but also that there are no two pixels that cover the same area in two consecutive frames taken by the same satellite. To calculate the \gls{gtfp}, let $T_o(h)$ be the orbital period of a satellite deployed at altitude $h$. Then, $T_o(h)$ can be closely approximated as 
\begin{equation}
    T_o(h)\approx \sqrt{\left(\frac{4\pi^2}{\text{G}M_E}\right)\left(R_E+h\right)^3},
    \label{eq:orb_period}
\end{equation}
where $\text{G}$ is the universal gravitational constant; $M_E$ and $R_E$ are the mass and radius of the Earth, respectively.
The \gls{gtfp}, denoted as $T_\text{GTF}$, is given as
\begin{equation}
T_\text{GTF}(h_\text{px},d_\text{gsd},h)= \frac{h_\text{px}d_\text{gsd}T_o(h)}{2\pi R_E}.
\end{equation} 
 We define $t_k=kT_\text{GTF}$ as the time when frame $k$ is captured by the satellite and, hence, $t_0=0$.

The frames may be processed before they are sent to ground, which involves, for example, correction of optical anomalies, compression, and encoding. Specifically, a frame captured by the source satellite $v_0$, can be processed either locally or in a distributed manner to reduce its size by a compression factor $\rho>1$, such that the resulting size of the frame is $D_k/\rho$.\\[-3em]

\nwtxt{\subsection{\Gls{smec} framework}
 In the following, we describe a general \gls{smec} framework for real-time and very-high resolution Earth observation where the task instructs a source satellite $v_0$ to capture $K$ consecutive frames. The frameworks is based on classical communication models for distributed processing architectures, but considers the distinctive characteristics of \gls{leo} satellite communications, such as limited bandwidth and links with heterogeneous capacity. 
\begin{enumerate}
    \item \emph{Segmentation}: At each time slot $k\in\set\{1,2,\dotsc,K\}$, where $K$ is the total number of frames in the task, the source satellite $v_0$ captures a frame $k$ with $W_k$ images for a total size $D_k$\,bits and divides it into segments.
    \item \emph{Allocation:} Each of the satellites in the \gls{smec} cluster for frame $k$, namely $n\in\set{N}_k$, will receive a segment, which is a fraction $x_k^{(n)}$ of the total amount of data in the frame. The \gls{gs} might also indicate a fraction of the frame $x_k^{(g)}$ that will not be processed by the satellites but downloaded directly without processing.
    \item \emph{Scatter}: The source satellite distributes the segments to all the satellites in the \gls{smec} cluster $n\in\set{N}_k$ using high-data rate \glspl{isl}. The rest of the data are routed directly to the \gls{gs} without processing. 
    \item \emph{Distributed processing}: The segments are processed at the satellites, with a consequent reduction in the size of the data by a factor $\rho_k$, called the compression ratio. Hence, the resulting size of the segment processed by satellite $n$ is $x_k^{(n)}/\rho_k$. The CPU frequency selected to process the $k$-th frame at satellite $n$ is denoted as $f_k^{(n)}$. 
    \item \emph{Gather}: Once processing is completed, the satellites in the \gls{smec} cluster $\set{N}_k$ send the processed segments to the \gls{gs} through the destination satellite $v_d$.
\end{enumerate}}

\subsection{Compression}
Let $C(\rho_k, \epsilon)$ be the complexity of the compression algorithm, defined as the number of CPU cycles to compress one bit of data by a compression ratio $\rho_k$, being $\epsilon$ a positive constant that depends on the compression algorithm. In traditional and pure JPEG compression, the complexity can be considered as constant w.r.t. the compression ratio $\rho_k$ since the latter is only determined by the entries of the quantization matrix, which does not affect the number of operations to be performed. Nevertheless, setting a constant compression complexity might only represent a small subset of possible compression algorithms.

Instead, we adopt a model where the complexity of the compression algorithm increases exponentially with the compression ratio as $C(\rho_k,\epsilon)=e^{\epsilon \rho_k}-e^{\epsilon}$~\cite{Li2019}. This latter model covers several popular compression techniques, such as Zlib, Zstandard, and XZ compression and makes our optimization framework more general. Consequently,  by selecting the exponential model for the complexity of the compression algorithm, our framework is applicable to a much wider range of compression algorithms than the constant complexity model for JPEG. Furthermore, as it will be seen in Section~\ref{sec:problem}, adoption of the exponential model creates a trade-off between the energy used for processing and for communication. In contrast, selecting the optimal compression ratio under a constant complexity model for the formulated optimization problem would be trivial. 

To process and compress the images, the processing payload of the satellites consists of a CPU with $N_\text{CPU}$ cores and a maximum clock frequency $f_\text{CPU}$. The satellites are able to adapt the CPU clock frequency for each frame $k\in\set{K}$ in the task through Dynamic Voltage and Frequency Scaling (DVFS), which allows to reduce the energy consumption of tasks with relaxed latency requirements. Therefore, we define $f_k^{(n)}$ as the CPU clock frequency selected by satellite $n\in\set{N}$ to process the data belonging to frame $k$. 

Let $T^\text{proc}_k$ be the execution time of the image processing algorithm for a frame in task $k$, which is given, for satellite $n$, as
\begin{equation}
T^\text{proc}_k(n,D_k,\rho,f_k;\epsilon, N_\text{CPU})=\frac{D_k\,C(\rho,\epsilon)}{N_\text{CPU}f_k^{(n)}}.
\label{eq:proc_time}
\end{equation}

The processing of the data received by the satellites is managed by a scheduler that operates in a \nwtxt{first}-in first-out fashion. Hence, it may need to queue the processing of a segment until the processing of the previous one is completed and the CPU is available. Specifically, we denote the queueing time to process the data received at time $t$ at satellite $n$ as $Q^\text{proc}_t(n)\in\mathbb{R}^+$\,seconds. In other words, the data received at time $t$ must wait in the queue of satellite $n$ until time $t+Q^\text{proc}_t(n)$ before it can be processed. Throughout this paper, we focus on maintaining the stability of the processing queues rather than on their impact on the overall latency.


\subsection{Communication}
After processing, the resulting data must be transmitted to the GS $g$. The communication modules are used to establish communication with the satellites in the same orbital plane via the intra-plane \glspl{isl}, and with $g$. Specifically, intra-plane \glspl{isl} are established with the two neighboring satellites in the same orbital plane and a downlink satellite-to-ground link is established between $g$ and the closest satellite. Let $\set{G}_k=\left(\set{V}, \set{E}_k\right)$ be the directed graph that represents the system at time $t_k$, where $\set{V}=\set{N}\cup\{g\}$ is the vertex set and $\set{E}_k$ is the edge set at time $t_k$. As mentioned above, we consider a time-slotted system and, hence, update graph $\set{G}_k$ along with its parameters by taking a snapshot at the beginning of each time slot.

Without loss of generality, we denote the source satellite as $v_0$ and the destination satellite, which has a direct connection to the \gls{gs}, as $v_d$. The directed edges $(u,v)\in\set{E}_k$ represent the full-duplex communication links, which are assumed to remain fixed within the interval $\left[t_k,t_{k+1}\right]$. A path from vertex $v_0$ to vertex $v_d$ is called a $v_0-v_d$ path and denoted as $\set{P}_k(v_0,v_d)=(v_0,v_1,\dotsc, v_d)$, where $v_0,v_1,\dotsc,v_d\in\set{V}$ and $\ell\left(\set{P}_k(u,v)\right)$ is the length of the path.

High-data rate \gls{fso} links are considered for the intra-plane \glspl{isl} since the relative distances and positions among these satellites are maintained, which minimizes pointing errors~\cite{Kaushal2017}. As a consequence, the intra-plane \glspl{isl} are fixed throughout the operation of the network and also possess a fixed data rate $R_\text{ISL}$ and \nwtxt{consume a fixed amount of power during transmission $P_\text{ISL}$.} 

Point-to-point satellite-to-ground links are established dynamically between the \gls{gs} and the closest available satellite. Therefore, we define the edge set at time $k$ as
\begin{equation}
    \set{E}_k=\left\{(u,v)\cup(v_d,g):u,v\in\set{N},v_d=\argmin_{v\in \set{N}} d_{t_k}(v,g)\right\},
\end{equation}
where $d_{t_k}(v,g)$ is the Euclidean distance between satellite $v$ and $g$ at time $t_k$. The latter can be accurately calculated based on the ephemeris of the satellites for any $t_k\in\mathbb{R}$ and, hence, our model operates with snapshots of the edge set taken at each  $t_k$.

Differently from the intra-plane ISLs, the beams and data rate must be adapted continuously at the downlink due to the rapid movement of the satellites. Traditional RF technology is less prone to outages due to pointing errors and atmospheric conditions than FSO and, hence, it is used to achieve reliable communication to the GS.  

We consider an interference-free additive-white Gaussian noise (AWGN) channel with free-space path loss and with noise power $\sigma^2$. Hence, the \gls{snr} for the satellite-to-ground link (i.e., downlink) at time $t$ is given as
\begin{equation}
    \gamma_t = G_\text{tx}\,G_\text{rx}\,P^\text{tx}_\text{RF}\left(\frac{\mathrm{c}}{4\pi\, d_t(v_d,g) \,f_c\,\sigma}\right)^2, \text{ where }  v_d=\argmin_{v} d_t(v,g)
\end{equation}
where $f_c$ is the carrier frequency, $\mathrm{c}=2.998\cdot10^8$~m/s is the speed of light, $G_d$ and $G_g$ are the transmitter and receiver antenna gains, and $P^\text{tx}_\text{RF}$ is the transmission power. 

\nwtxt{Once the \gls{snr} is known, a proper modulation and coding scheme can be selected as follows. Let $\mathcal{R}_\text{DVB}=\left\{R'\right\}$ be the set of available rates $R'$ in b/s/Hz defined in the DVB-S2X system. Next, let $\gamma_\text{min}(R')\geq 2^{R'}-1$ be the minimum required \gls{snr} to achieve a block error rate $<10^{-5}$ with rate $R'$. Then, the modulation and coding scheme for downlink communication is selected to achieve the data rate
\begin{equation}
    R_k=B\max \left\{R'\in\set{R}_\text{DVB-S2}:\min\{\gamma_{t}\}\geq \gamma_\text{min}(R'),\right.\,
    t\in \left[t_k,t_k+\Delta t\right]\}.\IEEEeqnarraynumspace
\end{equation}
That is, the rate is selected using the thresholds defined in the DVB-S2X system~\cite{dvb_s2} and $R'$ is a valid rate for downlink communication at time slot $k$ if and only if the \gls{snr} for this link is greater than $\gamma_\text{min}(R')$ throughout the whole time slot.}

Similarly as for processing, we denote the queueing time at the communication module to transmit the data from $u$ to $v$ as $Q^\text{trans}_t(u,v)\in\mathbb{R}^+$\,seconds, where $t$ is the time at which the data is received by $u$. Furthermore, we define the transmission time for a packet of size $D_k$ as
\begin{equation}
    L_k^\text{trans}(D_k,u,v)=\frac{D_k}{R_k(u,v)},
\end{equation}
where $R_k(u,v)=R_\text{ISL}$ for all the ISLs and $R_k(v_d,g)=R_k$ for the downlink.

\subsection{Energy consumption}
\nwtxt{In the considered problem, there are two contributors of interest to the energy consumption at the satellite: the processing and compression of the images, and the communications.}

We consider a model for the energy consumption of the satellites due to the processing of the task that captures the most relevant CPU parameters \cite{Cheng2018, Zhang2013, Mao2017}. In this model, the energy consumption of the CPU per clock cycle is proportional to the square of its clock frequency $f_k^{(n)}$ times a constant $\nu$, which is the effective capacitance coefficient~\cite{Mao2017, Zhang2013}. Specifically,
\begin{equation}
    E_\text{cycle}^\text{proc}\left(f_k^{(n)}\right) = \nu \, \left(f_k^{(n)}\right)^2=\frac{P_\text{proc}(f_\text{CPU})}{f_\text{CPU}^3} \, \left(f_k^{(n)}\right)^2,
\end{equation}
where $P_\text{proc}\left(f_\text{CPU}\right)$ is the power consumption during processing at the maximum CPU frequency.
Building on this, and assuming that the supplied power is linear with the number of processor cores $N_\text{CPU}$, the energy consumption to process an image of size $D_k$ bits is modeled as
\begin{equation}
    E^\text{proc}_{k}\left(\rho_k,f_k^{(n)};D_k,\epsilon\right) = D_k\, C(\rho_k,\epsilon)\, E_\text{cycle}^\text{proc}\left(f_k^{(n)}\right).
\label{eq:energy_proc}
\end{equation}

\nwtxt{Regarding the energy consumption during communication, we consider a model that includes the energy due to data transmission, the inefficiency of the power amplifiers, and the static power consumption of the communication modules~\cite{Matthiesen2020}. Consequently, the power consumption for the RF downlink during communication $P_\text{RF}$ is modeled as a function of the power consumption due to data transmission $P_\text{RF}^\text{tx}$, inefficiency of the power amplifier $\mu^\text{amp}_\text{RF}$, and the static power consumption $P_\text{RF}^\text{static}$, as
\begin{equation}
P_\text{RF}=\mu^\text{amp}_\text{RF}\, P_\text{RF}^\text{tx}+P_\text{RF}^\text{static}.
\label{eq:rf_power}
\end{equation}
We assume that the RF link of the destination satellite $v_d$ consumes $P_\text{RF}^\text{static}$ at all times and, hence, is not considered to evaluate the energy consumption of the \gls{smec} framework. Then, during data transmission, the $v_d$ consumes an extra $\mu^\text{amp}_\text{RF}\, P_\text{RF}^\text{tx}$~W and the energy consumption due to transmit $D_k$~bits of data in the downlink is 
\begin{equation}
    E_\text{RF}^\text{trans}(D_k)=\frac{\mu^\text{amp}_\text{RF}\, P_\text{RF}^\text{tx}\,D_k}{R_k}
\end{equation}}

\nwtxt{Similarly, the power consumption for the \gls{fso} \glspl{isl} during communication can be divided into the power consumption due to data transmission $P_\text{ISL}^\text{tx}$, the inefficiency of the power amplifier $\mu_\text{amp}$, and the static power consumption $P_\text{ISL}^\text{static}$, and is given as~\cite{Matthiesen2020}
\begin{equation}
P_\text{ISL}=\mu^\text{amp}_\text{ISL} \, P_\text{ISL}^\text{tx}+P_\text{ISL}^\text{static}.
\end{equation}
 In our model for the \gls{fso} links, we consider that $P_\text{ISL}$ is fixed and define the parameter 
 \begin{equation}
     \eta=\frac{\mu^\text{amp}_\text{ISL}\, P_\text{ISL}^\text{tx}}{\mu^\text{amp}_\text{ISL}\, P_\text{ISL}^\text{tx}+P_\text{ISL}^\text{static}}\in\left[0,1\right].
 \end{equation}
 Therefore, we consider that $\eta\, P_\text{ISL}=\mu^\text{amp}_\text{ISL} \, P_\text{ISL}^\text{tx}$ is consumed during data transmission~\cite{Matthiesen2020}. The remaining $(1-\eta)P_\text{ISL}=P_\text{ISL}^\text{static}$ is the static power consumption of the  \gls{fso} \glspl{isl}.  Hence, the energy consumption to transmit $D_k$~bits of data through an \gls{fso} \gls{isl} is
\begin{equation}
    E_\text{ISL}^\text{trans}(D_k) = \frac{\eta\, P_\text{ISL}\,D_k}{R_\text{ISL}}.
\end{equation} }\\[-4em]

\section{Problem formulation}
\label{sec:problem}
A task might consist of one or multiple frames, namely $K$, that are captured continuously by a source satellite $v_0$. Each of the $k$ frames contains $W_k\in\{0,1,\dotsc,\}$ images. The GS schedules the segmentation, processing, and transmission of the segments in a task jointly using its knowledge about the communication and processing capabilities of the satellites, along with the value of $W_k$ for all $k\in\set{K}$. For this, the GS builds an allocation matrix $\mathbf{X}\in\mathbb{N}^{K\times N+1}$, whose $(k,n)$-th element is $x_k^{(n)}$, indicating the amount of data from frame $k$ that satellite $n$ must process. Moreover, we denote the $k$-th row of $\mathbf{X}$ as
\begin{equation}
\mathbf{x}_k = \left[x_k^{(1)}, x_k^{(2)}, \dotsc, x_k^{(N)}, x_k^{(g)}\right],
\end{equation} 
which is the allocation vector for frame $k$.
Naturally, 
\begin{equation}
    \mathbf{x}_k\bm{1}=\sum_{n=1}^Nx_k^{(n)}+x_k^{(g)}=D_k
\end{equation}
and the number of segments for frame $k$ is the number of non-zero elements in vector $\mathbf{x}_k$.

The segment allocation is accompanied by matrix $\mathbf{F}\in\mathbb{R}_+^{K\times N}$, whose $(k,n)$-th element is $f_k^{(n)}$ and defines the CPU frequency that must be selected by the satellite $n$ for processing the segment from frame $k$. Finally, the GS must define the compression factor for each frame, defined by the vector $\bm{\rho} = \left[\rho_1,\rho_2,\dotsc, \rho_K\right]$. 

We follow a resource slicing approach to ensure the stability of the processing and communication queues in the orbital plane. In such approach, the task is allocated an amount of processing and communication resources that is proportional to its duration $KT_\text{GTF}$. Therefore, the processing and communication parameters for the task  values must be selected to ensure that the average processing and transmission time of each segment, at each satellite and each link does not exceed the GTFP. Building on this, we formulate the following constraints.

\emph{Processing constraint:} To ensure the stability of the processing queues at the satellite CPUs, the average time to process all the segments allocated to a given satellite $n\in\set{N}$, of size $\sum_{k=0}^{K}x_k^{(n)}$, must be shorter than the duration of the task. Therefore, a proper CPU frequency must be selected for the processing of each segment at each satellite, denoted as $f_k^{(n)}$.
From~\eqref{eq:proc_time} we derive the processing constraint: 
\begin{IEEEeqnarray}{C}
\sum_{k=1}^K \frac{x_k^{(n)}\,C(\rho_k,\epsilon)}{N_\text{CPU}\, f_k^{(n)}}\leq  K\,T_\text{GTF}, \quad \forall n\in\set{N}.\IEEEeqnarraynumspace
    \label{eq:c_proc_time}
\end{IEEEeqnarray}

\emph{Downlink and ISL rate constraints:} The average data rate for the satellite-to-ground link and the \glspl{isl} must be sufficiently high to transmit the generated data within the considered period of time of $K\,T_\text{GTF}$\,seconds. Hence, we formulate the downlink constraint as follows.
\begin{equation}
  \sum_{k=1}^K \left(x_k^{(g)}+\frac{1}{\rho_k}\sum_{n=1}^N x_k^{(n)}\right) \leq T_\text{GTF}(h_\text{px},d_\text{gsd},h)\sum_{k=1}^K R_{k}
    \label{eq:c_s2g_rate}
\end{equation}
For the ISLs, the total amount of traffic depends on the scatter algorithm and the location of the processing satellites $n\in\set{N}_k$ w.r.t. the source satellite $v_0$ and $g$ for all the $K$ frames. Specifically, to calculate the amount of traffic  assigned to each \gls{isl} $(u,v)\in\set{E}_k$, let $\set{P}_k(u,v)$ be the shortest path between $u$ and $v$ at time slot $k$. Next, we define the indicator variable $y_k^{(e)}(u,v)$, which takes the value of $1$ if the edge $e\in\set{E}_k$ is in the path $\set{P}_k(u,v)$ and $0$ otherwise. That is,
\begin{equation}
    y_k^{(e)}(u,v)\triangleq\begin{cases}
    1, & \text{if } e\in\set{P}_k(u,v) \\
    0 & \text{otherwise}.
    \end{cases}
\end{equation}  
Based on the latter, we define the constraint for the data rate at the ISLs as
\begin{equation}
    \sum_{k=1}^{K} x_k^{(g)}y_k^{(e)}(v_0,g)+\sum_{n=1}^N x_k^{(n)}\left(y_k^{(e)}\left(v_0,n\right)+\frac{y_k^{(e)}\left(n,g\right)}{\rho_k}\right)
    \leq  KT_\text{GTF}R_\text{ISL}, \quad\forall e\in \set{E}_k. 
     \label{eq:c_isl_rate}
\end{equation}
Nevertheless, in most practical scatter algorithms, the \glspl{isl} that communicate the source satellite $v_0$ with its neighbors will be subject to the heaviest traffic. Therefore, the latter constraint can be simplified by limiting it to the ISLs $\set{E}_0=\{(v_0,n)\in\set{E}:n\in\set{N}_k^{(n)}\}$

Further, recall that $\ell\left(\set{P}_k(u,v)\right)$ is defined as the length of the $u-v$ path. Then, we define the energy devoted for the \emph{scatter} phase as
\begin{IEEEeqnarray}{rl}
    E_k^\text{scatter}&(v_0, \mathbf{x}_k;\eta)=\nwtxt{\frac{\mu^\text{amp}_\text{RF}\, P^\text{tx}_\text{RF}\,x_k^{(g)}}{R_k}} +\frac{\eta P_\text{ISL}}{R_\text{ISL}}
   \left(\sum_{n=1}^N\ell\left(\set{P}_k(v_0,n)\right)x_k^{(n)} + \left(\ell\left(\set{P}_k(v_0,g)\right)-1\right)x_k^{(g)}\right)\!,\IEEEeqnarraynumspace
   \label{eq:e_scatter}
\end{IEEEeqnarray}
where the first addend corresponds to energy needed to reach the GS and the second one for the distribution among satellites. The parenthesis accounts for the amount of data that reaches each node, weighted by the number of steps (i.e., links). Likewise, the consumed energy for the \emph{gather} phase is defined as 
\begin{IEEEeqnarray*}{rl}
    E_k^\text{gather}&(\rho_k, \mathbf{x}_k;\eta)
    =\frac{1}{\rho_k}\sum_{n=1}^Nx_k^{(n)}\left(\nwtxt{\frac{\mu^\text{amp}_\text{RF}\,P^\text{tx}_\text{RF}}{R_k}}+\frac{\eta\, P_\text{ISL}}{R_\text{ISL}}\left(\ell\left(\set{P}_k(n,g)\right)-1\right)\right),\IEEEeqnarraynumspace\IEEEyesnumber
     \label{eq:e_gather}
\end{IEEEeqnarray*}
where each compressed segment, i.e., $x_k^{(n)}/\rho_k$, is weighted by the energy per bit it takes to reach the GS from each satellite $n$ through ISL and RF links. The overall energy consumption, due to distribution and processing of the data, is defined as the sum of the energy at the scatter, processing, and gather phases. Hence, we formulate the energy minimization problem as
    \begin{IEEEeqnarray*}{rrCll}
    \mathrm{P}_1:\,&\minimize_{\mathbf{X},\mathbf{F}, \bm{\rho}}& \quad &\IEEEeqnarraymulticol{2}{l}{\sum_{k=1}^KE_k^\text{scatter}(v_0,\mathbf{x}_k;\eta)+E_k^\text{gather}(\rho_k,\mathbf{x}_k;\eta)+\sum_{n=1}^N E_k^\text{proc}\left(x_k^{(n)},\rho_k,f_k^{(n)};\epsilon\right)}
    \IEEEyesnumber\label{eq:global_parallel_processing}\\
&\text{subject to} &\quad &\eqref{eq:c_proc_time},~\eqref{eq:c_s2g_rate},~\eqref{eq:c_isl_rate}\\
&&&0\leq x_k^{(n)}\leq D_k,\qquad&\forall n\in\set{N}, k\in\set{K}\IEEEyessubnumber\IEEEeqnarraynumspace\\
&&&\mathbf{x}_k\bm{1}=D_k,\quad&\forall k\in\set{K}\IEEEyessubnumber\IEEEyessubnumber\IEEEeqnarraynumspace\\
&& &1<\rho_k\leq\rho_\text{max},&\forall k\in\set{K}\IEEEyessubnumber\label{const:rho}\\
&&&0<f_k^{(n)}\leq f_\text{CPU},&\forall k\in\set{K}, n\in\set{N}. \IEEEyessubnumber
\end{IEEEeqnarray*}
where the parameter $\rho_\text{max}$ is a pre-defined maximum acceptable compression ratio.

That is, for each frame $k\in\set{K}$, the GS must determine the compression ratio  $\rho_k$, the allocation vector $\mathbf{x}_k$, and the CPU frequency $f_k^{(n)}$ to minimize energy while fulfilling the processing and communication constraints. \nwtxt{Note that constraint~\eqref{const:rho}} is defined to avoid the value of $\rho_k=1$, as this represents no processing at the satellites and is an equivalent solution to setting $x_k^{(g)}=1$.

Naturally, the routing algorithm, which determines the paths to be followed by the segments, has an impact on the links used during the scatter and gather phases by determining the variables $\ell\left(\set{P}_k(v_0,n)\right)$, $\ell\left(\set{P}_k(n,g)\right)$,  and $y_k^{(e)}(v_0,n)$ and $y_k^{(e)}(n,g)$.
Consequently, the routing algorithm has an impact on the ISL constraint and on the energy consumption. Nevertheless, investigating the optimal routing algorithm for each phase is out of the scope of the paper and, hence, we consider a typical hop-count shortest-path routing algorithm and treat these variables as parameters in~\eqref{eq:global_parallel_processing}.

\nwtxt{For the case with}  $C(\rho_k,\epsilon)=e^{\rho_k\epsilon}-e^\epsilon$, neither the energy consumption for processing $E_k^\text{proc}\left(\rho_k, f_k^{(n)};D_k,\epsilon\right)$, defined in~\eqref{eq:energy_proc}, nor for communication in the gather phase $E_k^\text{gather}(\rho_k,\mathbf{X}_k;\eta)$, defined in~\eqref{eq:e_gather} are jointly convex in $\mathbf{X}$ and $\bm{\rho}$.
Consequently, the objective function~\eqref{eq:global_parallel_processing} is not jointly convex in $\mathbf{X}$, $\bm{\rho}$, and $\mathbf{F}$. Furthermore, the constraints \eqref{eq:c_proc_time},~\eqref{eq:c_s2g_rate},~\eqref{eq:c_isl_rate} are not jointly convex in $\mathbf{X}$ and $\bm{\rho}$. Thus, $\text{P}_1$ is a non-convex optimization problem. Nevertheless, we can exploit the fact that the problem is convex when only considering one optimization variable at a time. In particular, a closed-form expression can be obtained for the CPU frequency of the satellites $\mathbf{F}$. Furthermore, $E_k^\text{scatter}(v_0,\mathbf{x}_k;\eta)$ is linear in $\mathbf{X}$.
Therefore, we follow a variable decomposition approach and decompose the general optimization problem~\eqref{eq:global_parallel_processing} into three sub-problems, which are solved iteratively following the \gls{bcd} algorithm to reach a near-optimal solution~\cite{Tseng2001}. Specifically, we begin by obtaining the closed-form solution for the optimal CPU frequency $\mathbf{F}^*$. The optimal CPU frequency is then used to optimize $\mathbf{X}$ and $\bm{\rho}$ iteratively.

Conversely, for the case with the JPEG compression algorithm $C(\rho_k,\epsilon)=\epsilon$, the optimal solution for the compression ratio is  $\rho_k=\rho_\text{max}$, which is treated as a parameter and the optimization problem~\eqref{eq:global_parallel_processing} simplifies to only optimizing $\mathbf{X}$ and $\mathbf{F}$ jointly.

\subsection{Optimizing the CPU frequency}
As starting point, we consider the problem of optimizing the CPU frequency $\mathbf{F}$ for a given $\bm{\rho}$ and $\mathbf{X}$ in $\text{P}_1$. We formulate the optimization problem for this case as
\begin{IEEEeqnarray*}{rrCll}
    \mathrm{P}_F:\,&\minimize_\mathbf{F}& ~& \sum_{n=1}^N\sum_{k=1}^KE_k^\text{proc} \left(x_k^{(n)},\rho_k,f_k^{(n)};\epsilon\right)  
    \IEEEyesnumber\label{eq:optimize_f}\\
&\text{subject to} &\quad &\sum_{k=1}^K \frac{x_k^{(n)}C(\rho_k,\epsilon)}{KN_\text{CPU}\, f_k^{(n)}}\leq  T_\text{GTF}~&\forall  n\in\set{N}\\
&&&0\leq f_k^{(n)}\leq f_\text{CPU},&\forall k\in\set{K}, n\in\set{N}. 
\end{IEEEeqnarray*}

\begin{theorem}[Optimal CPU frequency] 
If the problem is feasible for the given $\mathbf{X}$ and $\mathbf{\rho}$, the optimal CPU frequency is constant and equals the minimum required to satisfy the processing constraint~\eqref{eq:c_proc_time}.
Therefore, the optimal CPU frequency for satellite $n$ given $\mathbf{X}$ and $\bm{\rho}$ is given by
\emph{\begin{equation}
    f_k^{(n)*} = f^{(n)}= \displaystyle\frac{\sum_{k=1}^Kx_k^{(n)}C(\rho_k,\epsilon)}{K\,N_\text{CPU}\,T_\text{GTF}}, \quad \forall n\in\set{N}:\displaystyle\sum_{k=1}^K x_k^{(n)}C(\rho_k,\epsilon)  -T_\text{GTF}K\,N_\text{CPU}\, f_\text{CPU}\leq0.
    \label{eq:opt_f}
\end{equation}}
\label{th:p_freq}
The proof is based on the complementary slackness condition and is presented in the Appendix~\ref{sec:appendix}.
\end{theorem}
In all the cases where the processing constraint cannot be fulfilled, the CPU frequency is set momentarily to $f^{(n)}=f_\text{CPU}$ to proceed with the iterative optimization.

\subsection{Optimizing the task allocation}
Next, we define problem $\mathrm{P}_\mathbf{X}$: optimizing $\mathbf{X}$ for a given $\mathbf{F}^*$ and $\bm{\rho}$. Based on the previous result for $\mathbf{F}^*$, we transform the constraint~\eqref{eq:c_proc_time} and define the problem as
\begin{IEEEeqnarray*}{rrCll}
    \mathrm{P}_\mathbf{X}:\,&\minimize_{\mathbf{X}}& \quad & \IEEEeqnarraymulticol{2}{l}{\sum_{k=1}^K \left(E_k^\text{scatter}(v_0,\mathbf{x}_k;\eta)+E_k^\text{gather}(\rho_k,\mathbf{x}_k;\eta)\right)+\sum_{n=1}^N E_k^\text{proc}\left(x_k^{(n)},\rho_k,f^{(n)};\epsilon\right)} \IEEEeqnarraynumspace
    \IEEEyesnumber\label{eq:optimize_x}\\
&\text{subject to} &\quad &\sum_{k=1}^K x_k^{(n)}\,C(\rho_k,\epsilon)-  K\,T_\text{GTF}\,N_\text{CPU}\, f^{(n)}\leq 0,\quad & \forall n\in\set{N},\IEEEyessubnumber\IEEEeqnarraynumspace
    \label{eq:c_proc_time_lag}\\
    &&&\eqref{eq:c_s2g_rate},~\eqref{eq:c_isl_rate}\\
&&&0\leq x_k^{(n)}\leq D_k,\quad&\forall n\in\set{N}, k\in\set{K}\IEEEeqnarraynumspace\IEEEyessubnumber\\
&&&\mathbf{x}_k\bm{1}=D_k,\quad&\forall k\in\set{K}\IEEEeqnarraynumspace\IEEEyessubnumber\\[-2em]
\end{IEEEeqnarray*}

\nwtxt{Note that $\mathrm{P}_\mathbf{X}$} is linear in $\mathbf{X}$ with affine constraints and, since $\mathbf{F}^*$ can be calculated in closed-form, we solve the problem using an iterative convex optimization approach with the augmented Lagrangian of~\eqref{eq:c_proc_time_lag}, where the processing constraint is moved to the objective and used as penalty~\cite{Bertsekas1976}. Specifically,
\begin{IEEEeqnarray*}{rrCll}
    \mathrm{P}_{\set{L}\left(\mathbf{X},\mathbf{F}^*;\alpha\right)}\!:\,&\minimize_{\mathbf{X}}& \!\phantom{a} & \IEEEeqnarraymulticol{2}{l}{\sum_{k=1}^K\left(E_k^\text{scatter}(v_0,\mathbf{x}_k;\eta)+E_k^\text{gather}(\rho_k,\mathbf{x}_k;\eta)\right)\!+\!\sum_{n=1}^N\Bigg[ E_k^\text{proc}\!\left(x_k^{(n)},\rho_k,f^{(n)};\epsilon\right)}\\
    &&&\IEEEeqnarraymulticol{2}{l}{+\lambda^{(n)}\!\left(\sum_{k=1}^K \frac{x_k^{(n)}C(\rho_k,\epsilon)}{  K\,T_\text{GTF}\,N_\text{CPU}}- f^{(n)}+s^{(n)}\right)}\\
    &&&\IEEEeqnarraymulticol{2}{l}{+\frac{\alpha}{2}\left(\sum_{k=1}^K \frac{x_k^{(n)}C(\rho_k,\epsilon)}{  K\,T_\text{GTF}\,N_\text{CPU} }-f^{(n)}+s^{(n)}\right)^2\Bigg]}\IEEEeqnarraynumspace
    \IEEEyesnumber\label{eq:optimize_x_lag}\\
&\text{subject to} &&
    \eqref{eq:c_s2g_rate},~\eqref{eq:c_isl_rate}\\
&&&0\leq x_k^{(n)}\leq D_k,\quad&\forall n\in\set{N}, k\in\set{K}\IEEEeqnarraynumspace\IEEEyessubnumber\\
&&&\mathbf{x}_k\bm{1}=D_k,\quad&\forall k\in\set{K}.\IEEEeqnarraynumspace\IEEEyessubnumber\\
\end{IEEEeqnarray*}

We solve problem~\eqref{eq:optimize_x_lag} following the \gls{ipdd} method \cite{Shi2020} by first solving for $\mathbf{X}$ for a given $\mathbf{F}$.
Then, the values of $\mathbf{F}^*$ are updated after each iteration with the update rule in~\eqref{eq:opt_f}. Finally, as $\mathbf{F}^*$ has been already updated, the slack variables $s^{(n)}$ become $0$. Then, following the \gls{ipdd} method, the Lagrange multipliers $\lambda^{(n)}$ and the penalty terms $\alpha^{(n)}$ are updated according to a pre-defined threshold $\tau_\text{proc}$ and increase factor $\beta\in(0,1)$. Namely, if the processing constraint is below a pre-defined threshold $\tau_\text{proc}$, the update rule for $\lambda^{(n)}$ with inequality constraints becomes
\begin{equation}
    \lambda^{(n)} = \max\left(0,\lambda^{(n)} + \alpha^{(n)}\left( \sum_{k=1}^Kx_k^{(n)}C(\rho_k,\epsilon) -T_\text{GTF}\,K\,N_\text{CPU}\,f^{(n)}\right)\right).
    \label{eq:lambda_n}
\end{equation}
Otherwise, if the processing constraint is above $\tau_\text{proc}$, the penalty parameter  is increased as $\alpha^{(n)}\leftarrow \alpha^{(n)}/\beta$.
\subsection{Optimizing the compression ratio}
Finally, we proceed to optimize the compression factor $\bm{\rho}$ for a given $\mathbf{X}$ and $\mathbf{F}^*$ for the case where $C(\rho_k,\epsilon)$ increases with $\rho_k$. 
The optimization problem for $\bm{\rho}$ is formulated from~\eqref{eq:global_parallel_processing} by removing the constant term $E_k^\text{scatter}(v_0,\mathbf{x}_k;\eta)$ and the constraints on $\mathbf{X}$, which give
\begin{IEEEeqnarray*}{rrCll}
    \mathrm{P}_{\bm{\rho}}:\,&\minimize_{\bm{\rho}}& \quad & \IEEEeqnarraymulticol{2}{l}{\sum_{k=1}^KE_k^\text{gather}(\rho_k,\mathbf{x}_k;\eta)+\sum_{n=1}^N E_k^\text{proc}\left(x_k^{(n)},\rho_k,f^{(n)};\epsilon\right)} \IEEEeqnarraynumspace
    \IEEEyesnumber\label{eq:optimize_rho}\\
&\text{subject to} &\quad &\sum_{k=1}^K x_k^{(n)}C(\rho_k,\epsilon)-  K\,T_\text{GTF}\,N_\text{CPU} \,f^{(n)}\leq 0,\quad & \forall n\in\set{N},\IEEEyessubnumber\IEEEeqnarraynumspace
\\
    &&&\eqref{eq:c_s2g_rate},~\eqref{eq:c_isl_rate}\\
    &&&1<\rho_k\leq \rho_\text{max} & \text{for all } k\in\set{K} \IEEEyessubnumber
\end{IEEEeqnarray*}

 Let $\bm{s}$ be the vector of slack variables, which includes those for the downlink constraint $s^{(g)}$, for the \gls{isl} constraints $s^{(e)}$, and for the values of $\rho_k$, namely $s_k^{(\rho)}$. Further, we define the Lagrange multiplier and penalty terms for these constraints as $\lambda^{(g)}$ and $\alpha^{(g)}$, $\lambda^{(e)}$ and $\alpha^{(e)}$, and $\lambda_k^{(\rho)}$ and $\alpha_k^{(\rho)}$, respectively. Building on these, the augmented Lagrangian for $ \mathrm{P}_{\bm{\rho}}$ is given by~\eqref{eq:aug_lag_rho}\addtocounter{equation}{1}
\setcounter{storeeqcounter}{\value{equation}} on top of Page~\pageref{eq:aug_lag_rho}.
 \begin{figure*}[!t]
  \normalsize
  \setcounter{tempeqcounter}{\value{equation}}
\begin{IEEEeqnarray*}{rl}
\IEEEeqnarraymulticol{2}{l}{\set{L}(\bm{\rho},\bm{\lambda}, \bm{s};\mathbf{X},\bm{\alpha})=}
\\
\qquad &\sum_{k=1}^{K}E_k^\text{gather}(\rho_k,\mathbf{x}_k;\eta)+ \sum_{n=1}^N\Bigg[E_k^\text{proc}\left(x_k^{(n)},\rho_k,f^{(n)};\epsilon\right)\\ 
&+\lambda^{(n)}\!\left( \sum_{k=1}^K\frac{x_k^{(n)}C(\rho_k,\epsilon)}{K\,N_\text{CPU}\,T_\text{GTF}} -f^{(n)}\!+s^{(n)}\!\right)\!+\frac{\alpha^{(n)}}{2}\left( \sum_{k=1}^K\frac{x_k^{(n)}C(\rho_k,\epsilon)}{K\,N_\text{CPU}\,T_\text{GTF}} -f^{(n)}\!+s^{(n)}\!\right)^{\!2}\Bigg]\\
&+\lambda^{(g)}\left[s^{(g)}+\sum_{k=1}^K\frac{x_k^{(g)}+\sum_{n=1}^N\frac{x_k^{(n)}}{\rho_k}}{T_\text{GTF}}-R_k\right]+\frac{\alpha^{(g)}}{2}\left[s^{(g)}+\sum_{k=1}^K\frac{x_k^{(g)}+\sum_{n=1}^N\frac{x_k^{(n)}}{\rho_k}}{T_\text{GTF}}-R_k\right]^2\\
&+\sum_{e\in\set{E}_s}\lambda^{(e)}\left[s^{(e)}+\sum_{k=1}^{K} \frac{x_k^{(g)}y_k^{(e)}(v_0,g)}{K\,T_\text{GTF}\,R_\text{ISL}}+\sum_{n=1}^N \frac{x_k^{(n)}\left(y_k^{(e)}\left(v_0,n\right)+\frac{y_k^{(e)}\left(n,g\right)}{\rho_k}\right)}{K\,T_\text{GTF}\,R_\text{ISL}}\right]\\
&+\sum_{e\in\set{E}_s}\frac{\alpha^{(e)}}{2}\left[s^{(e)}+\sum_{k=1}^{K} \frac{x_k^{(g)}y_k^{(e)}(v_0,g)}{K\,T_\text{GTF}\,R_\text{ISL}}+\sum_{n=1}^N \frac{x_k^{(n)}\left(y_k^{(e)}\left(v_0,n\right)+\frac{y_k^{(e)}\left(n,g\right)}{\rho_k}\right)}{K\,T_\text{GTF}\,R_\text{ISL}}\right]^2\\
&+\sum_{k=1}^{K}\lambda^{(\rho)}_k\left(\rho_k-\rho_\text{max}+s_k^{(\rho)}\right)+\sum_{k=1}^{K}\frac{\alpha^{(\rho)}_k}{2}\left(\rho_k-\rho_\text{max}+s_k^{(\rho)}\right)^2 \IEEEyesnumber
\label{eq:aug_lag_rho}
\end{IEEEeqnarray*}
  \hrulefill
  \vspace*{-1em}
\end{figure*}\setcounter{equation}{\value{tempeqcounter}+1}
Note that neither the slack variables for the downlink constraint nor for the \gls{isl} constraints depend on $k$ as the limiting factors are the average rates throughout the execution of the task. Therefore, the problem becomes
\begin{equation}
    \minimize_{\bm{\rho}\geq 1,\, \bm{s}\geq 0}~ \maximize_{\bm{\lambda}\geq0,\, \lambda^{(n)},\, \lambda_k} \,\set{L}(\mathbf{X},\bm{\rho},\bm{\lambda}, \bm{s};\bm{\alpha}),
\end{equation}
which we solve using the method of projected gradient descent by applying the following update rules for $\bm{\lambda}$ after updating $\mathbf{F}$ according to~\eqref{eq:opt_f} and $\lambda^{(n)}$ according to~\eqref{eq:lambda_n}.

\begin{IEEEeqnarray}{rCl}
    \lambda^{(g)} &=& \max\Bigg(0,\lambda^{(g)} + \alpha^{(g)}\Bigg[s^{(g)}+\sum_{k=1}^K\frac{1}{T_\text{GTF}}\left(x_k^{(g)}+\sum_{n=1}^N\frac{x_k^{(n)}}{\rho_k}\right)-R_k\Bigg]\Bigg)
\end{IEEEeqnarray}

\begin{IEEEeqnarray}{rCl}
    \lambda^{(e)} &=& \max\Bigg(0,\lambda^{(e)} + \alpha^{(e)}\left[s^{(e)}+\sum_{k=1}^{K} \frac{x_k^{(g)}y_k^{(e)}(v_0,g)}{K\,T_\text{GTF}\,R_\text{ISL}}+\sum_{n=1}^N \frac{x_k^{(n)}\left(y_k^{(e)}\left(v_0,n\right)+\frac{y_k^{(e)}\left(n,g\right)}{\rho_k}\right)}{K\,T_\text{GTF}\,R_\text{ISL}}\right]\IEEEeqnarraynumspace
    \end{IEEEeqnarray}
    
\begin{equation}
     \lambda_{k}^{(\rho)} = \max\left(0,\lambda_{k}^{(\rho)}+\alpha^{(\rho)}_k\left(\rho_k-\rho_\text{max}-s^{(\rho)}_k\right)\right)
\end{equation}

\subsection{\nwtxt{Global optimization algorithm and practical considerations}}
Algorithm~\ref{alg:iterative} illustrates the iterative procedure for global optimization for the case where $C(\rho_k,\epsilon)=e^{\rho_k\epsilon}-e^\epsilon$. Note that a feasible value 
for each $\rho_k$ must be selected during initialization. For instance, we initialize each $\rho_k$ with the value
\begin{equation}
    \rho_k^\text{init} = \min\left\{\max\left(1, \frac{D_k}{T_\text{GTF}\,R_k}\right), \rho_\text{max}\right\}.
    \label{eq:rho_min}
\end{equation}
\nwtxt{As with traditional penalty methods, the convergence of the \gls{ipdd} method and of projected gradient descent depend on the initial value of the penalty terms and the termination condition. While these methods can achieve convergence with finite penalty terms and finite number of iterations, the number of iterations to fulfill the termination conditions vary widely depending on the characteristics of the objective function and the problem constraints. In our study, the termination condition for the optimization of the task allocation $\mathbf{X}$ is $\|\mathbf{X}-\mathbf{X}'\|_2\leq \delta$ and of the compression ratio $\bm{\rho}$ is  $\|\bm{\rho}-\bm{\rho}'\|_2\leq \delta$. These termination conditions are defined in lines $5$ and $10$ of Algorithm \ref{alg:iterative}. If Algorithm \ref{alg:iterative} were to be implemented in a real system where the solution must be obtained within a specific deadline, the termination conditions can be modified to limit the number of iterations so the deadline can be met. This will result in a sub-optimal solution, but the execution time can be easily characterized for the specific computing platform.
}

\begin{algorithm} [t]
	\centering
	\caption{\gls{bcd} algorithm for iterative non-convex optimization.}
	\begin{algorithmic}[1] 
	\renewcommand{\algorithmicrequire}{\textbf{Input:}}
		\renewcommand{\algorithmicensure}{\textbf{Output:}}
		\REQUIRE $K$, $R_\text{ISL}$ $W_k$, $\set{G}_k$, $R_k$ for all $k\in\set{K}$ and tolerance $\delta$
		\STATE  Calculate $\set{P}_k(v_0,n)$, $y_k^{(e)}(v_0,n)$,  $\set{P}_k(n,g)$ and $y_k^{(e)}(n,g)$ for all $n\in\set{N}$ and $k\in\set{K}$
		\STATE Initialize $x_k^{(v_0)}\leftarrow D_k$, $x_k^{(n)}=x_k^{(g)}\leftarrow0$ for all $n\neq v_0$, and $\rho_k\leftarrow\rho_k^\text{init}$ for all $k\in\set{K}$, and $f^{(n)}\leftarrow f_\text{CPU}$ for all $n$
		\STATE $\mathbf{X}'\leftarrow \bm{0}$ and $\bm{\rho}'\leftarrow \bm{0}$
		\WHILE {$\|\mathbf{X}-\mathbf{X}'\|_2+\|\bm{\rho}-\bm{\rho}'\|_2>\delta$}
		\WHILE  {$\|\mathbf{X}-\mathbf{X}'\|_2>\delta$}
		\STATE $\mathbf{X}'\leftarrow \mathbf{X}$
		\STATE Optimize $\mathbf{X}$ given $\mathbf{F}^*$ and $\bm{\rho}$ with \gls{ipdd}~\eqref{eq:optimize_x_lag}
		\STATE Update $\mathbf{F}^*$ given $\mathbf{X}$ and $\bm{\rho}$ as in~\eqref{eq:opt_f}
		\ENDWHILE
		\WHILE  {$\|\bm{\rho}-\bm{\rho}'\|_2>\delta$}
		\STATE $\bm{\rho}'\leftarrow \bm{\rho}$
		\STATE Optimize $\bm{\rho}$ given $\mathbf{F}^*$ and $\mathbf{X}$ with projected gradient descent as in~\eqref{eq:aug_lag_rho}
		\STATE Update $\mathbf{F}^*$ given $\mathbf{X}$ and $\bm{\rho}$ as in~\eqref{eq:opt_f}
		\ENDWHILE
		\ENDWHILE
		\RETURN $\mathbf{X}^*\leftarrow \mathbf{X}$, $\rho^*\leftarrow \bm{\rho}$, and $\mathbf{F}^*$
\end{algorithmic} 
\label{alg:iterative}
\end{algorithm}


\section{Results}
\label{sec:results}
We consider an orbital plane with $N=20$ satellites capturing images in an HD format with $1920\times 1080$ pixels under the following two scenarios.
\begin{itemize}
    \item\emph{Per-frame optimization:} Optimizing one frame independently is optimal if the task is relatively long (i.e., $K\to\infty$) and the frame size is constant $W_k=W_{k'}$ for all $k, k'\in\set{K}$ and, naturally, if $K=1$. 
    \item\emph{Multi-frame optimization:} In all cases, optimizing across the $K$ frames of the task is the optimal approach. We illustrate the gains of this approach with respect to per-frame optimization in a synthetic scenario that leads to the upper bound in energy savings and in a realistic scenario that leads to the lower bound in energy savings.
\end{itemize}
Without loss of generality, the source satellite is denoted as $v_0$. The destination satellite $v_d$ is assumed to be located at the edge of coverage at the first frame of the task and moves towards the center of coverage. The default parameters for the performance analysis are listed in Table~\ref{tab:params}. With these parameters, the downlink data rate at the edge of coverage area is $2.16$\,Gbps.
The \gls{gsd} is set to $d_\text{gsd}=0.5$\,m, which results in a \gls{gtfp} of $78$\,ms. 

The results were obtained using a simulator coded in Python to replicate the orbital dynamics of the satellites and to calculate the data rates at each time slot. The optimization problems were solved using the CVXPY package~\cite{diamond2016cvxpy} using MOSEK ApS as solver.


\begin{table}[t]
    \centering
    \caption{Parameter \nwtxt{settings for performance evaluation}.}
    \begin{tabular}{@{}llll@{}}
        \toprule 
        Parameter &  Symbol & Setting\\\midrule
         Altitude of deployment $[\mathrm{km}]$& $h$ & $600$\\
          Number of satellites & $N$ & $20$\\
          Processor frequency  $[\mathrm{GHz}]$ & $f_\text{CPU}$ & $1.8$\\
          Number of processor cores & $N_\text{cores}$ & $4$\\
          Power consumption for processing $[\mathrm{W}]$ & $P_\text{proc}(f_\text{CPU})$& $10$\\
          Data rate of the ISLs $[\mathrm{Gbps}]$ & $R_\text{ISL}$ & $10$\\
          Transmission power of the ISLs $[\mathrm{W}]$ & $P_\text{ISL}$ & $60$\\
          Downlink transmission power $[\mathrm{W}]$ & $P^\text{tx}_\text{RF}$ & $10$\\
          \nwtxt{Inefficiency of the downlink RF power amplifier} & \nwtxt{$\mu^\text{amp}_\text{RF}$} & \nwtxt{$1$}\\
          Downlink carrier frequency $[\mathrm{GHz}]$ & $f_c$ & $20$\\
          Downlink bandwidth $[\mathrm{MHz}]$& $B$ & $500$\\
          Downlink antenna gain $[\mathrm{dBi}]$ & $G_d$ & $32.13$\\
         Antenna gain of the GS $[\mathrm{dBi}]$ & $G_g$ & $34.20$\\
         \nwtxt{Noise power $[\mathrm{dBW}]$} & \nwtxt{$\sigma_\text{dB}^2$} & \nwtxt{$-119.32$}\\ 
          Width of the $k$-th frame $[\mathrm{images}]$ & $W_k$ & $\{0,1,2,\dotsc\}$\\ 
          Size of HD image $[\mathrm{MB}]$ & $D_\text{img}$ & $ 5.93$\\
          Ground sample distance (GSD) $[\mathrm{m/pixel}]$ & $d_\text{gsd}$ & $0.5$\\ 
        Maximum compression factor &$\rho_\text{max}$ & $20$\\
         Complexity of the image processing algorithm & $\epsilon$ & $0.1$\\
         \nwtxt{Fraction of $P_\text{ISL}$ consumed during data transmission} &$\eta$ & $\{0.1,1\}$\\
         \bottomrule
    \end{tabular}
    \label{tab:params}
\end{table}

\subsection{Per-frame optimization}
\begin{figure}
    \centering
    \subfloat[]{\includegraphics{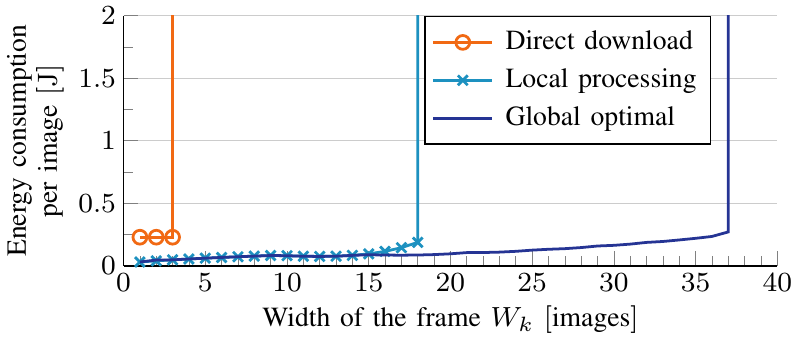} \label{fig:econs_ell1_eta01}}\subfloat[]{\includegraphics{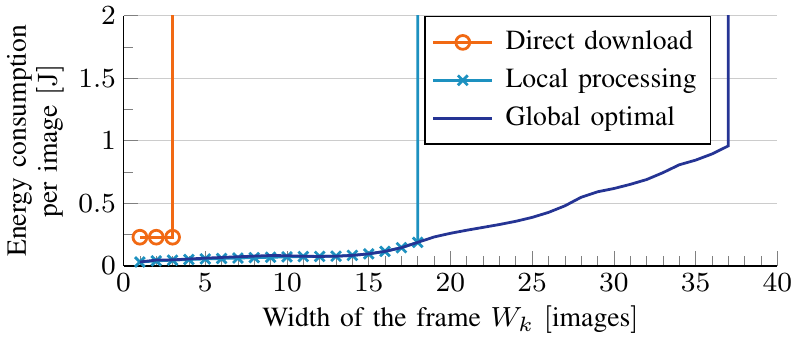} \label{fig:econs_ell1}}\\
    \subfloat[]{\includegraphics{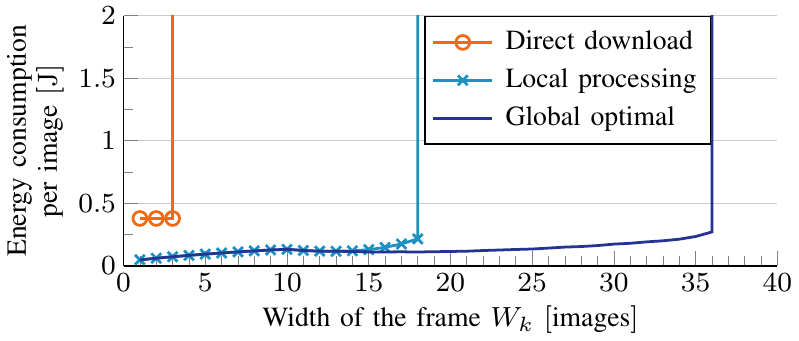}\label{fig:econs_ell6_eta01}}\subfloat[]{\includegraphics{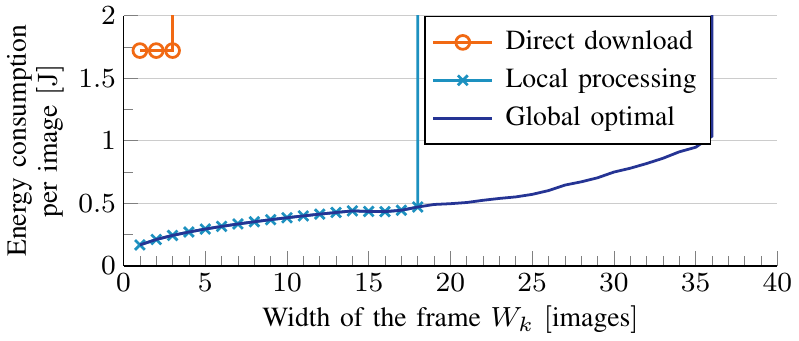}\label{fig:econs_ell6}}
    \caption{Energy consumption per image for $d_\text{gsd}=0.5$\,m with per-frame optimization for the cases where the destination satellite is $v_d=v_0$ (a) with $\eta=0.1$ and (b) with $\eta=1$ and where the destination satellite is $v_d=v_5$ (c) with $\eta=0.1$ and (d) with $\eta=1$ considering the three approaches: direct download, local processing, and  global optimal. The energy consumption is set to $\infty$ outside the feasible region.}
    \label{fig:econs_ell}
\end{figure}

Fig.~\ref{fig:econs_ell} shows the energy consumption per image for the feasible values of $W_k$ for direct download, local processing, and the global optimal solution. Two topologies are considered: 1) in Fig.~\ref{fig:econs_ell1} the destination satellite is the same as the source satellite $v_d=v_0$  and 2) in Fig.~\ref{fig:econs_ell6} the destination satellite is $v_d=v_5$, namely, it's five hops away from the source satellite. Besides, two values are considered for parameter $\eta=\{0.1,1\}$.

As it can be seen in Fig.~\ref{fig:econs_ell}, the energy consumption with direct download is much higher when compared to local processing and to the global optimal. Furthermore, direct download is only feasible with $W_k\leq 3$ images. \nwtxt{In contrast, local processing achieves a similar energy consumption as the global optimal for most of the cases within its feasible region $W_k\leq18$ images. This behavior means that local processing is the optimal solution with relatively small frames sizes $W_k$.} Finally, the global optimal solution leads to the minimum energy consumption in all cases and also increases the feasible region to $W_k\leq37$ for $v_d=v_0$ (see Fig.~\ref{fig:econs_ell1_eta01} and Fig.~\ref{fig:econs_ell1}) and to $W_k\leq 36$ for $v_d=v_5$ (see Fig.~\ref{fig:econs_ell6_eta01} and Fig.~\ref{fig:econs_ell6}). Thus, the global optimal solution increases the number of images per frame that are supported by the system by a factor of $12\times$ when compared to direct download and by a factor of $2\times$ when compared to local processing.

\nwtxt{In addition, the energy consumption per image is considerably lower for $v_d=v_0$ in Fig.~\ref{fig:econs_ell1} than for $v_d=v_5$ in Fig.~\ref{fig:econs_ell6} due to the increased distance between the source satellite and the destination, which increases the energy consumption at the gather phase.} Naturally, the energy consumption with $\eta=0.1$ (see Fig.~\ref{fig:econs_ell1_eta01} and Fig.~\ref{fig:econs_ell6_eta01}) is considerably lower than that with $\eta=1$ (see Fig.~\ref{fig:econs_ell1} and Fig.~\ref{fig:econs_ell6}).

\begin{figure}[t]
    \centering
    \subfloat[]{\includegraphics{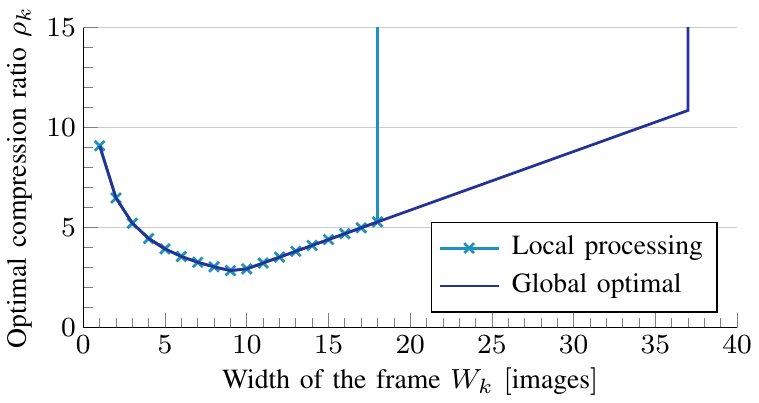}}\hfil
    \subfloat[]{\includegraphics{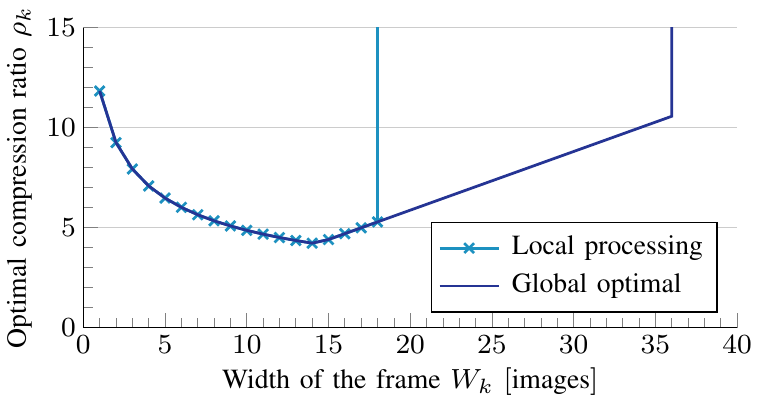}}
    \caption{Optimal compression ratio with per-frame optimization considering local processing and global optimal for (a) $v_d=v_0$ and  (b) $v_d=v_5$.}
    \label{fig:rho}
\end{figure}

Next, Fig.~\ref{fig:rho} shows the selected value of $\rho$ as a function of $W_k$. Clearly, the same value of $\rho_k=\rho_k^*$ is selected by both the global optimal and the local processing approaches. Specifically, $\rho_k^*$ decreases as $W_k$ increases up to $W_k=9$ for $v_d=v_0$ and up to $W_k=14$ for $v_d=v_5$. \nwtxt{As $W_k$ increases beyond these values, the optimal compression ratio is $\rho_k^*=D_k/R_k$, which is the lowest value that ensures that the downlink constraint is fulfilled, which can be easily calculated in closed form.}

\begin{figure}[t]
    \centering
    \subfloat[]{\includegraphics{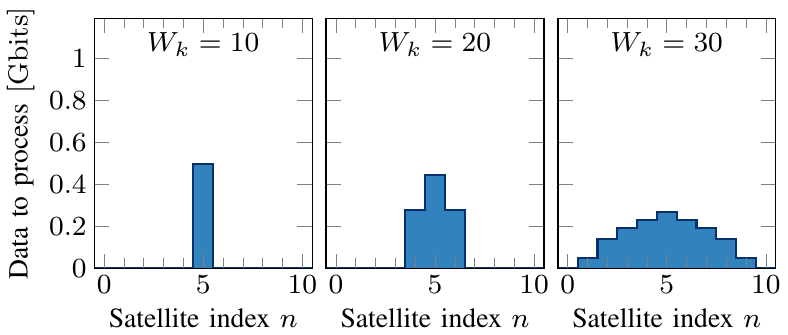}\label{fig:X_per_sat_v0_e01}}\hfil
    \subfloat[]{\includegraphics{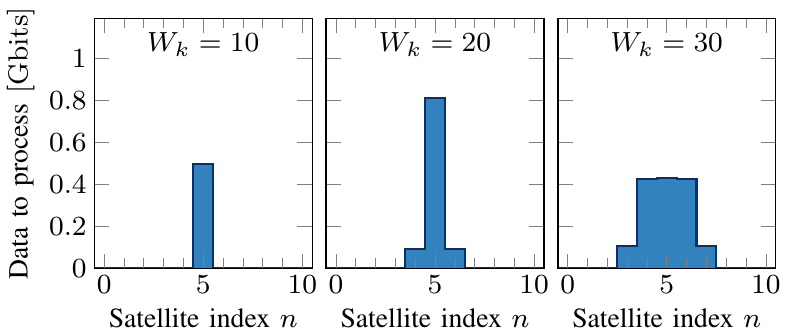}\label{fig:X_per_sat_v0_e1}}\\[-0.8em]
    \subfloat[]{\includegraphics{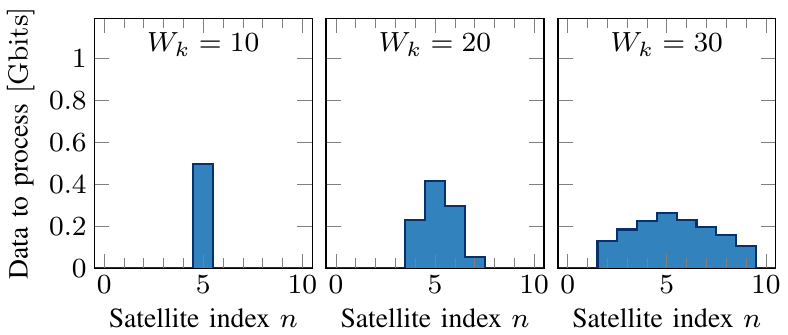}\label{fig:X_per_sat_v5_e01}}\hfil
    \subfloat[]{\includegraphics{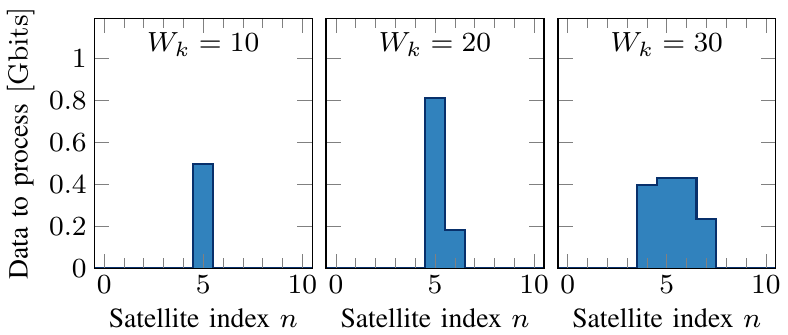}\label{fig:X_per_sat_v5_e1}}
    \caption{\nwtxt{Amount of data processed at the \gls{smec} cluster for each satellite $n$ for $W_k=\{10,20,30\}$ with the source satellite $v_0$ being $n=5$. The destination satellite $v_d$ is also $n=5$ in (a) $\eta=0.1$ and  (b) $\eta=1$ and the destination satellite $v_d$ is $n=10$ in (c) $\eta=0.1$  and (d) $\eta=1$.}}
    \label{fig:X_per_sat}
\end{figure}

We conclude the analyses of the per-frame optimization scenario by illustrating the values of $x_k^{(n)*}$ in Fig.~\ref{fig:X_per_sat} for $v_d = v_0$ and $v_d=v_5$ with $\eta=\{0.1,1\}$ and the source satellite being $n=5$. Clearly, Fig.~\ref{fig:X_per_sat} shows that the data to process at satellites close to the source $n=5$ increases as $W_k$ increases. Specifically, as shown in Fig.~\ref{fig:X_per_sat_v0_e01} and Fig.~\ref{fig:X_per_sat_v0_e1}, the data is distributed symmetrically across neighbouring satellites in the \gls{smec} cluster when $v_d=v_0$ as the source satellite has direct connection to the \gls{gs}. Conversely, for the case with $v_d=v_5$ shown in Fig.~\ref{fig:X_per_sat_v5_e01} and Fig.~\ref{fig:X_per_sat_v5_e1}, a larger amount of the data is distributed to satellites in the \gls{smec} cluster that are in the shortest path towards the destination $n=10$. \nwtxt{Furthermore, more satellites are used for processing for the case with $\eta=0.1$ (see Fig.~\ref{fig:X_per_sat_v0_e01} and Fig.~\ref{fig:X_per_sat_v5_e01}) when compared to the case with $\eta=1$ (see Fig.~\ref{fig:X_per_sat_v0_e1} and Fig.~\ref{fig:X_per_sat_v5_e1}). This is because the case with $\eta=1$ results in a higher transmission power in the \gls{fso} links, which serves as a penalty for inter-satellite communication. In all the illustrated cases, the amount of data downloaded without processing is zero. }

\subsection{Multi-frame optimization}
Next, we evaluate the energy consumption that can be achieved by performing multi-frame optimization. First, we consider a scenario where the amount of tasks per \gls{gtfp} is relatively low and the scheduler can dedicate an extended duration per task to reduce energy consumption. This is modeled by defining an initial frame of width $W_0$ images followed by $K-1$ empty frames (i.e., with $W_k=0$ for $k>0$). 

\begin{figure}[t]
    \centering
    \subfloat[]{\includegraphics{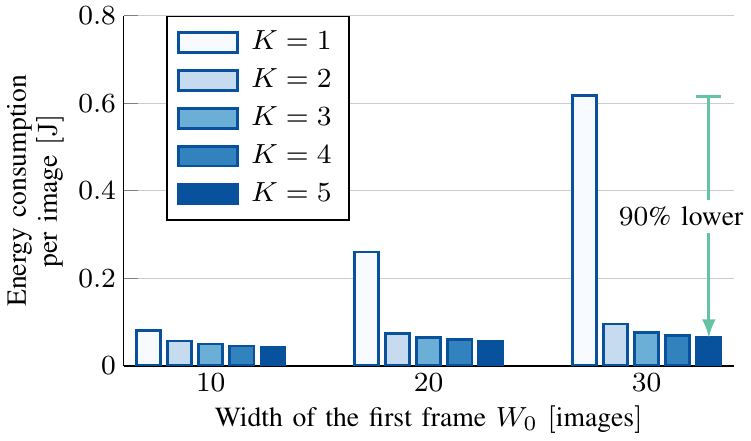}\label{fig:empty_frames_v0}}\hfil
    \subfloat[]{\includegraphics{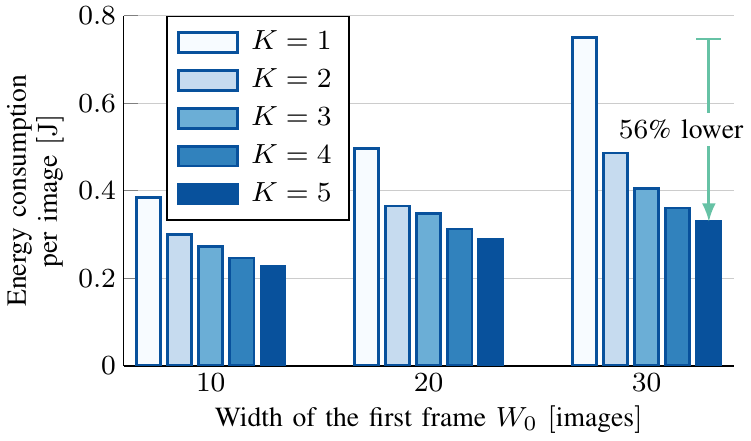}\label{fig:empty_frames_v5}}
    \caption{Global optimal energy consumption for a task with $K$ frames for $\eta=1$, where the first frame contains $W_0\in\{5,10,15,20\}$ images and the remaining $K-1$ frames are empty, i.e., $W_k=0$ for $k>0$. (a) $v_d=v_0$ and (b) $v_d=v_5$.}
    \label{fig:empty_frames}
\end{figure}

The energy consumption for $W_0=\{10,20,30\}$ with $K=\{1,2,3,4,5\}$ is shown in Fig.~\ref{fig:empty_frames} considering $v_d=v_0$ in Fig.~\ref{fig:empty_frames_v0} and $v_d=v_5$ in Fig.~\ref{fig:empty_frames_v5}. As it can be seen, the energy consumption increases drastically with the width of the frame $W_0$ for the case with $K=1$, which has zero empty frames. On the other hand, it is clear that the energy consumption per image decreases as $K$ increases. Specifically, the energy consumption with $K=5$, which contains $4$ empty frames, is $90$\% lower than with $K=1$ for the case with $v_d=v_0$ and $56$\% lower than with $K=1$ for the case with $v_d=v_5$. 

\begin{figure}[t]
\subfloat[]{\includegraphics{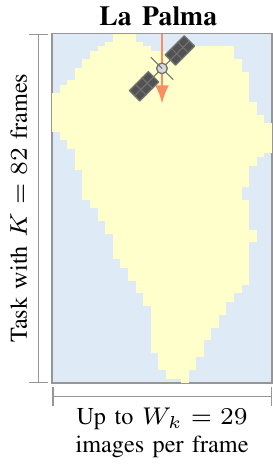}\label{fig:LaPalma_map}}\hfil
\subfloat[]{\includegraphics{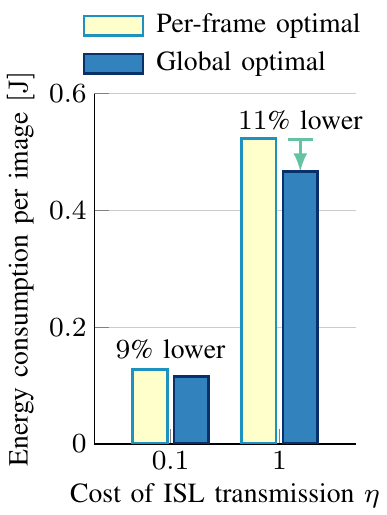}\label{fig:LaPalma_e_cons}}
\subfloat[]{\includegraphics{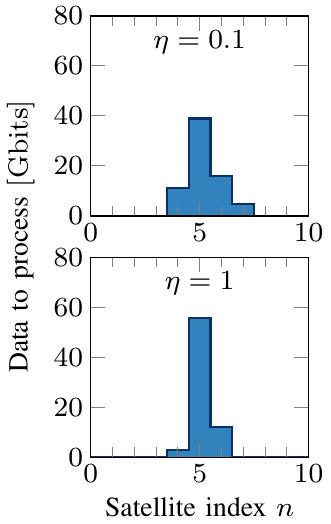}\label{fig:LaPalma_alloc}}
\subfloat[]{\includegraphics{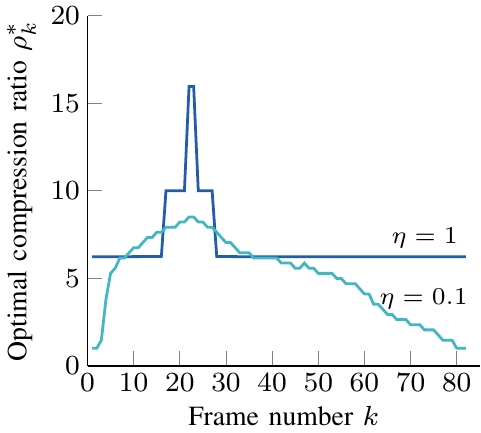}\label{fig:LaPalma_rho}}
\caption{(a) Task characteristics, (b) energy consumption, \nwtxt{(c) task allocation where the source satellite $v_0$ is $n=5$}, and (d) optimal value of $\rho_k$ for scanning La Palma island with a \gls{gsd} of $d_\text{gsd}=0.5$\,m.}
\label{fig:LaPalmaResults}
\end{figure}

We conclude our analyses by evaluating the energy savings for the case of a realistic task of scanning the island of La Palma with $v_d=v_5$. Such task comprises $K=82$ frames with different widths $W_k\in[1,29]$ as shown in Fig.~\ref{fig:LaPalma_map}. The total duration of the task is equal to the duration of the pass of the satellite over La Palma takes $6.4$\,s, occurs from top to bottom of the area shown in Fig.~\ref{fig:LaPalma_map}. Due to the large number of images per frame, neither direct download nor local processing are feasible in this scenario.  

Fig.~\ref{fig:LaPalma_e_cons} shows the energy consumption per image with per-frame optimization and with global optimization for $\eta=\{0.1,1\}$ for scanning the island of La Palma. 
Naturally, the energy consumption with $\eta=0.1$ is much lower than with $\eta=1$ as the energy consumption for transmission at the \glspl{isl} is $10\times$ greater for the latter case. Most importantly, Fig.~\ref{fig:LaPalma_e_cons} shows that the global optimization across all the $K=82$ frames results in saving $9$\% and $11$\% of the energy when compared to per-frame optimization for $\eta=0.1$ and $\eta=1$, respectively. Even though these energy savings are lesser than for the case with $K-1$ empty frames, they still represent an important reduction in energy consumption.

\nwtxt{Next, Fig.~\ref{fig:LaPalma_alloc} shows the amount of data allocated to each satellite throughout the $K=82$ frames $\sum_{k=1}^K x_k^{(n)}$ given that the source satellite is $n=5$. Clearly, the data is more evenly distributed across $4$ satellites with $\eta=0.1$ than with $\eta=1$, where the source satellite $n=5$ processes more than $78$\% of the total amount of data.} 

Finally, Fig.~\ref{fig:LaPalma_rho} shows that the optimal compression ratio for each frame $\rho_k^*$ is not uniform across the task and, in general, increases with the number of images in the frame. Furthermore, the values of $\rho_k^*$ are widely different for the two considered values of $\eta$, which illustrates the need for the careful selection of $\rho_k^*$ based on the task allocation and the transmission power.

\section{Conclusions}
\label{sec:conclusions}
In this paper, we considered a scenario where the physical characteristics of the system, dictated by the orbital parameters of the satellites and the area covered by their camera, determine the real-time requirements, which are necessary to maintain stability in the communication links and in the CPUs of the satellites. Moreover, we presented a general framework and an iterative optimization method for distributed \gls{smec}. Our results show that distributed \gls{smec} allows to capture, process, and download up to $6\times$ more images than with direct download and up to $2\times$ more images than with local \gls{smec}. Furthermore, up to $90$\% of the energy can be saved by carefully selecting the allocation of data, the compression ratio, and the processing frequency at the satellites. Finally, we considered a real-life task and observed that 1) it is only feasible to complete the task with distributed \gls{smec} and 2) optimizing the task parameters jointly leads to additional energy savings when compared to optimizing each frame independently. These benefits set the basis for achieving very-high resolution Earth observation missions.

\appendix
\label{sec:appendix}
To obtain the closed-form expression for the optimal CPU frequency, observe that the energy for processing increases monotonically with $f_k^{(n)}$. Furthermore, the problem is feasible with a given $\mathbf{X}$ and $\bm{\rho}$ if and only if $\exists F:0\leq f_k^{(n)}\leq f_\text{CPU}$ for all $n$ and $k$ that satisfies~\eqref{eq:c_proc_time}. 

To find the closed-form expression for the optimal CPU frequency, we define the Lagrangian of  $\mathrm{P}_F$, defined in~\eqref{eq:optimize_f} as
\begin{IEEEeqnarray} {rCl}
    \set{L}(\mathbf{F},\bm{\lambda})&=&\sum_{n=1}^N\Bigg(\sum_{k=1}^K \nu C(\rho_k,\epsilon)x_k^{(n)}\left(f_k^{(n)}\right)^2+\lambda^{(n)}\left(\sum_{k=1}^K \frac{x_k^{(n)}C(\rho_k,\epsilon)}{KN_\text{CPU} f_k^{(n)}}-  T_\text{GTF}\right)\nonumber\\
    &&+\sum_{k=1}^K\lambda_k^{(n)}\left(f_k^{(n)}-f_\text{CPU}\right)\Bigg)
\end{IEEEeqnarray}
From complementary slackness~\cite{Boyd}, we have that an optimal solution to the primal and dual problem satisfies the following two conditions.
\begin{equation}
\lambda^{(n)}\left(\sum_{k=1}^K \frac{x_k^{(n)}C(\rho_k,\epsilon)}{KN_\text{CPU} f_k^{(n)}}-  T_\text{GTF}\right)=0,
\label{eq:comp_slack_1}
\end{equation}
\begin{equation}
    \sum_{n=1}^N\sum_{k=1}^K\lambda_k^{(n)*}\left(f_k^{(n)*}-f_\text{CPU}\right)= 0.
    \label{eq:comp_slack_2}
\end{equation}

Thus, we express the first complementary slackness condition~\eqref{eq:comp_slack_1} as
\begin{equation}
\sum_{k=1}^K \frac{x_k^{(n)}C(\rho_k,\epsilon)}{ f_k^{(n)}}<  T_\text{GTF}KN_\text{CPU}\quad\Longrightarrow \quad\lambda^{(n)}=0,
\label{eq:slackness_1a}
\end{equation}
otherwise
\begin{equation}
\lambda^{(n)}>0  \quad\Longrightarrow \quad   \sum_{k=1}^K \frac{x_k^{(n)}C(\rho_k,\epsilon)}{ f_k^{(n)}}=  T_\text{GTF}KN_\text{CPU}.
\label{eq:slackness_1b}
\end{equation}
For condition~\eqref{eq:comp_slack_2} we have that
\begin{equation}
    f_k^{(n)*}<f_\text{CPU}\quad\Longrightarrow \quad\lambda_k^{(n)*}=0.
    \label{eq:slackness_2a}
\end{equation}
Otherwise,
\begin{equation}
    \lambda_k^{(n)*}>0\quad\Longrightarrow \quad f_k^{(n)*}=f_\text{CPU}.
    \label{eq:slackness_2b}
\end{equation}
Next, by taking the gradient of the Lagrangian $\set{L}(\mathbf{F},\bm{\lambda})$ w.r.t. $f_k^{(n)}$ we have
\begin{IEEEeqnarray}{rCl}
    \nabla_{f_k^{(n)}} \set{L}(\mathbf{F},\bm{\lambda})&=&
     2\nu C(\rho_k,\epsilon)x_k^{(n)}f_k^{(n)} - \frac{\lambda^{(n)}x_k^{(n)}C(\rho_k,\epsilon)}{KN_\text{CPU}\left(f_k^{(n)}\right)^2}+\lambda_k^{(n)} = 0.
     \label{eq:gradient}
\end{IEEEeqnarray}
In the following, we consider the solution of problem $P_F$ for the cases where $C(\rho_k,\epsilon)x_k^{(n)}>0$.

\textbf{Case 1:} To fulfill condition~\eqref{eq:slackness_2a}, the optimal value of the multiplier obtained from~\eqref{eq:gradient} is
\begin{equation}
    \lambda^{(n)*} = 2\nu KN_\text{CPU}\left(f_k^{(n)*}\right)^3, \qquad \text{s.t. } f_k^{(n)*}<f_\text{CPU} \text{ and } \lambda_k^{(n)*}=0 \text{ for all } k\in\set{K},
    \label{eq:opt_lam}
\end{equation}
Therefore, we conclude that an equal solution $f^{(n)}=f_k^{(n)*}<f_\text{CPU}$ is obtained for all $k\in\set{K}$. That is, for any given satellite $n\in\set{N}$, the optimal CPU frequency is equal for all the frames in a task. Furthermore, this implies that, if $\exists\lambda_k^{(n)}=0$ for $k\in\set{K}$, then $\lambda_k^{(n)}=0$ for all $k\in\set{K}$.

Specifically, by substituting $f_k^{(n)*}$ with $f^{(n)}$ in~\eqref{eq:slackness_1b}, we obtain
\begin{equation}
    f^{(n)} = \frac{1}{KN_\text{CPU}T_\text{GTF}}\sum_{k=1}^Kx_k^{(n)}C(\rho_k,\epsilon)
    \label{eq:opt_f_n}
\end{equation}

Furthermore, note that from~\eqref{eq:slackness_1b} and~\eqref{eq:opt_lam}, the following condition must hold
\begin{equation}
    f^{(n)}\in\left(0,f_\text{CPU}\right)\quad \Longrightarrow \lambda^{(n)*}>0.
\end{equation}  
Nevertheless, the latter does not prevent the case where $f^{(n)}=f_\text{CPU}$ and $\lambda^{(n)*}=0$ and, therefore~\eqref{eq:opt_f_n} is the optimal solution for all cases where $\lambda_k^{(n)}=0$.

\textbf{Case 2:}  To fulfill condition~\eqref{eq:slackness_1a}, the optimal value of the multiplier obtained from~\eqref{eq:gradient} is
\begin{equation}
    \lambda_k^{(n)*}=-2\nu C(\rho_k,\epsilon)x_k^{(n)}f_k^{(n)*}
    \label{eq:lam_k}
\end{equation}
Since all the terms on the right-hand side of~\eqref{eq:lam_k} are either positive or zero and it is required that $\lambda^{(n)*}\geq0$, then the only solutions to fulfill~\eqref{eq:lam_k} lead to $\lambda^{(n)*}=\lambda_k^{(n)*}=0$ and are as follows.

If there is no data to process at satellite $n$, any CPU frequency can be selected, namely,
\begin{equation}
C(\rho_k,\epsilon)x_k^{(n)}=0\quad\Longrightarrow \quad f_k^{(n)*}\in\left[0,f_\text{CPU}\right],
\end{equation}
which means that the value of $f_k^{(n)}$ is irrelevant. 

The other option is when the satellite $n$ has data to process
\begin{equation}
C(\rho_k,\epsilon)x_k^{(n)}>0\quad\Longrightarrow \quad f_k^{(n)*}=0,
\end{equation}
but the latter option makes the processing time to be infinite, so it should be excluded from the set of possible solutions. Consequently, conditions~\eqref{eq:slackness_1a} and~\eqref{eq:slackness_2a} can only be fulfilled jointly if the satellite has no data to process. In other words, there is no optimal solution where $f_k^{(n)*}<f_\text{CPU}$ and that leads to a processing time lesser than $T_\text{GTF}$.
\bibliographystyle{IEEEtran}
\bibliography{bib}
\end{document}